\documentclass[english,aps,reprint,superscriptaddress]{revtex4-2}
\usepackage{graphicx}
\usepackage{subfigure}
\usepackage{hyperref}
\usepackage{soul}
\usepackage{physics}
\usepackage{floatrow}
\usepackage{amssymb}
\usepackage{xfrac}
\usepackage{amsmath}
\usepackage[dvipsnames]{xcolor}
\usepackage[framemethod=TikZ]{mdframed}

\begin{document}

\title{Quantum control of molecules for fundamental physics}

\author{D. Mitra}
\affiliation{Department of Physics, Columbia University, 538 West 120th Street, New York, NY 10027-5255, USA}
\author{K. H. Leung}
\affiliation{Department of Physics, Columbia University, 538 West 120th Street, New York, NY 10027-5255, USA}
\author{T. Zelevinsky}
\email{tanya.zelevinsky@columbia.edu}
\affiliation{Department of Physics, Columbia University, 538 West 120th Street, New York, NY 10027-5255, USA}
\affiliation{Niels Bohr Institute, University of Copenhagen, Blegdamsvej 17, DK-2100 Copenhagen, Denmark}

\begin{abstract}
The extraordinary success in laser cooling, trapping, and coherent manipulation of atoms has energized the efforts in extending this exquisite control to molecules. Not only are molecules ubiquitous in nature, but the control of their quantum states offers unparalleled access to fundamental constants and possible physics beyond the Standard Model.  Quantum state manipulation of molecules can enable high-precision measurements including tests of fundamental symmetries and searches for new particles and fields. At the same time, their complex internal structure presents experimental challenges to overcome in order to gain sensitivity to new physics.  In this Perspective, we review recent developments in this thriving new field.  Moreover, throughout the text we discuss many current and future research directions that have the potential to place molecules at the forefront of fundamental science.
\end{abstract}

\maketitle

\section{Introduction}

The manipulation of pure quantum states of atoms and molecules is central to the advancement of quantum technology and quantum sensors. At the same time, low-energy high-precision measurements with atoms and molecules present a viable and scalable alternative to high-energy particle colliders in searches for physics beyond the Standard Model (SM). While currently a bedrock of our understanding of fundamental physical interactions, none of the SM particles can adequately account for cosmological observations of the energy and matter content, or the matter-antimatter asymmetry, in the Universe. From the perspective of fundamental science, the driving impetus for gaining deterministic control over atoms and molecules lies in the high-energy physical phenomena probed by their spectra measured with utmost precision. Broadly speaking, low-energy high-precision measurements enable the following \cite{SafronovaRMP18_NewPhysicsAtomsMolecules}: 
\begin{enumerate}
  \item Tests of quantum electrodynamics (QED), and observations of parity symmetry ($P$) violation as a manifestation of the electroweak interaction.
  \item Searches for time-reversal symmetry ($T$) violation through the observation of permanent electric dipole moments (EDMs) in elementary particles such as electrons, protons and neutrons, or nuclear magnetic quadrupole moments in composite systems with total spin greater than $1/2$.
  \item Searches for more exotic physics such as dark matter, ``fifth" forces, extra dimensions, and spatio-temporal variation of fundamental constants.
\end{enumerate} 

Molecules are ubiquitous in chemical and biological processes, yet efforts to achieve full quantum state control have concentrated extensively on atoms over the past 40 years \cite{ChuNature02_ColdAtomsQuantumControl}.  The variety of molecular species grows exponentially with the number of constituent atoms, and molecules possess vibrational and rotational degrees of freedom that atoms do not have. These traits broaden the tool set available to experimentalists, allowing a chosen molecule to be tailored for specific tasks. A series of molecules exhibit attractive properties for tests of fundamental physics that make them competitive with or superior to atoms. Very light molecules, such as molecular hydrogen isotopologues, are excellent systems for testing state-of-the-art QED calculations. Conversely, by combining high-precision spectroscopy with QED, it is possible to improve the precision of certain fundamental quantities such as the electron-to-proton mass ratio ($\eta\equiv m_e/m_p$), or to set constraints on yet unknown forces. A permanent EDM of a fundamental particle would directly imply $T$ violation (and therefore charge conjugation parity symmetry ($CP$) violation, assuming that $CPT$ is a good symmetry). The effect of a permanent EDM on molecular spectra is greatly enhanced in polar molecules or in deformed nuclei $-$ and especially in a combination of both. At energy scales close to the Planck energy ($\sim 10^{16}$ TeV), possible grand unification of the forces can involve violations of the Lorentz and $CPT$ symmetries, and lead to spatio-temporal variations of fundamental constants that would leave an imprint on molecular spectra. 

In this Perspective, we give an overview of the exciting developments in controlling individual quantum states of molecules, with an emphasis on fundamental physics applications. The astonishing progress in recent decades may be seen in a broader context as part of the ``second quantum revolution", where attention is shifted from studying states of matter that are inherently quantum mechanical (such as superconductivity) to engineering designer quantum states that offer advantages for certain tasks \cite{GeorgescuNoriRMP14}. 
We note that the challenges in molecular quantum state control are shared with the field of quantum simulation that seeks to model, for example, many-body spin systems where pseudo-spins could be encoded into the rotational structure of molecules.  Molecules in superpositions of their rotational states possess long-range, tunable dipolar interactions that are well suited for exploring this paradigm. Lanthanide atoms, such as Er and Dy, have been successfully cooled and trapped, and possess permanent magnetic dipole moments; however, these are typically weaker than the permanent electric dipole moments of heteronuclear diatomic molecules by one to two orders of magnitude \cite{Chomaz_arXiv_2022_magnetic_atoms_review}. This extensive research area is beyond the scope of this paper. Instead, we point the reader to numerous reviews on this quickly developing subject \cite{GeorgescuNoriRMP14,YeCarrNJP09_ColdMolecules,CornishBlackmoreQST19_RotationalCoherence_CaF_RbCs,DiVincenzoBennettNature00_QI,DeMillePRL02}, and note that refined control of molecular quantum states is also a route toward a deeper understanding of cold and ultracold chemical processes \cite{SoftleyHeazlewoodNChemRev21_ColdChemistry,NiHuScience19_KRbReactions}.

This paper is structured as follows.  Section \ref{sec:MolStructure} is pedagogical and introduces the concepts in molecular structure that are particularly important for understanding the approaches to quantum state control.  In Sec. \ref{sec:QuantumControlMolecules} we discuss modern methods for molecular state control, including techniques to create cold and ultracold molecular ensembles via control of motion, and manipulation of vibrational, rotational, and spin states.  Section \ref{Sec:FundSym} describes current and future applications of molecules in tests of fundamental symmetries.  In Sec. \ref{Sec:QEDNewForces} we present the status and expected development of molecular experiments designed for probing the subtleties of QED and attaining sensitivity to possible new forces that may exist in nature.  Finally, Sec. \ref{Sec:Future} provides a preview of additional future directions in this field, and Sec. \ref{Sec:Conclusion} concludes the Perspective.

\section{Molecular structure}
\label{sec:MolStructure}

The internal structure of molecules underlies their utility for high-precision measurements. It combines well separated quantum states with a reduced symmetry compared to atoms, leading to many opportunities in quantum control.

Molecules, unlike atoms, possess vibrational and rotational structure in addition to their electronic structure. The energy scales cover several orders of magnitude from $>100$~THz (electronic) to $\sim$10~THz (vibration), $\sim$10~GHz (rotation), and $\sim$10~MHz (hyperfine). Under the Born-Oppenheimer approximation, the electronic degrees of freedom are effectively decoupled from the rovibrational motion, and electronic states are described using the following operators and quantum numbers:

\begin{itemize}
    \item{Total orbital angular momentum of the electrons is given by the operator $\Vec{L}$. The projection of this operator along the internuclear axis is $L_z$ and \textbf{$\Lambda$} is its eigenvalue;}
    \item{Total spin angular momentum of the electrons is given by the operator $\Vec{S}$ and its projection $\Sigma$;}
    \item{Total angular momentum of the molecule is given by $\Vec{J}$ and its projections are given by \textbf{$\Omega$} and $M$ in the molecule-fixed and space-fixed frames, respectively.} 
    \item{For molecules with nuclear spin $\Vec{I} \ne 0$, the total angular momentum is given by $\Vec{F} = \Vec{I}+\Vec{J}$.}
\end{itemize}

Next, molecular states can be described according to the degree of coupling between the total orbital angular momentum $\Vec{N} = \Vec{L} + \Vec{R}$ (here $\Vec{R}$ is the rigid-body rotation operator) and the electronic degrees of freedom into so-called Hund's cases. The three principal cases for molecules discussed in this Perspective are:

\begin{enumerate}
\setlength{\itemsep}{0pt}
    \item{Hund's case (a): Strong coupling of the orbital angular momentum to the internuclear axis, and to the electronic spin (spin-orbit coupling). States like \textbf{$^{2S+1}\Pi_\Omega$} and \textbf{$^{2S+1}\Delta_\Omega$} fall under this case. The multiplicity of such a state is given by $\Omega = |\Lambda-\Sigma|,...,(\Lambda+\Sigma)$. 
    Each state with $\abs{\Lambda} \ne 0$ can take a $\pm\Lambda$ value. This leads to a degeneracy in each rotational level within the electronic manifold, which is known as $\Lambda$-doubling. The splitting between the doublets increases with higher $N$.}
    \item{Hund's case (b): Strong coupling of the orbital angular momentum to the internuclear axis, and to the rotational angular momentum in states like \textbf{$^{2S+1}\Sigma_\Omega$}. Although $\Lambda$-doubling does not occur in this case, the reflection symmetry about any plane passing through both nuclei of a diatomic molecule leads to two separate states denoted by $\Sigma^+$ and $\Sigma^-$. The latter is usually much higher in energy. As a result of the interaction between $N$ and $S$, each rotational level within the electronic manifold is split into $2S+1$ or $2N+1$ sublevels, whichever is smaller, by a spin-rotation interaction $\gamma\vec{N} \cdot \vec{S}$. The value of $\gamma$ can be relatively small, $\sim100$ MHz, for ground $\Sigma$ states. Spin-rotation coupling plays a crucial role in rotational closure during photon cycling (discussed in Sec. \ref{SubSec:direct_cooling} and Fig. \ref{fig:laser_cooling}(a)).}
    \item{Hund's case (c): Particularly for the case of heavy constituent atoms, the interaction between spin and orbital angular momenta can be much stronger than any coupling to rotational angular momentum, {and to the internuclear axis}. Then $\Lambda$ and {$\Sigma$} are not defined while $\Omega$ is a good quantum number, and each electronic state is split into a doublet known as $\Omega$-doubling, again borne out of the degeneracy of states defined by $\pm\Omega$. Examples include the $H$ state in ThO, and the $(1)0_u^+$ state in Sr$_2$.}
\end{enumerate}

\begin{figure}[!]
\includegraphics[width=\textwidth]{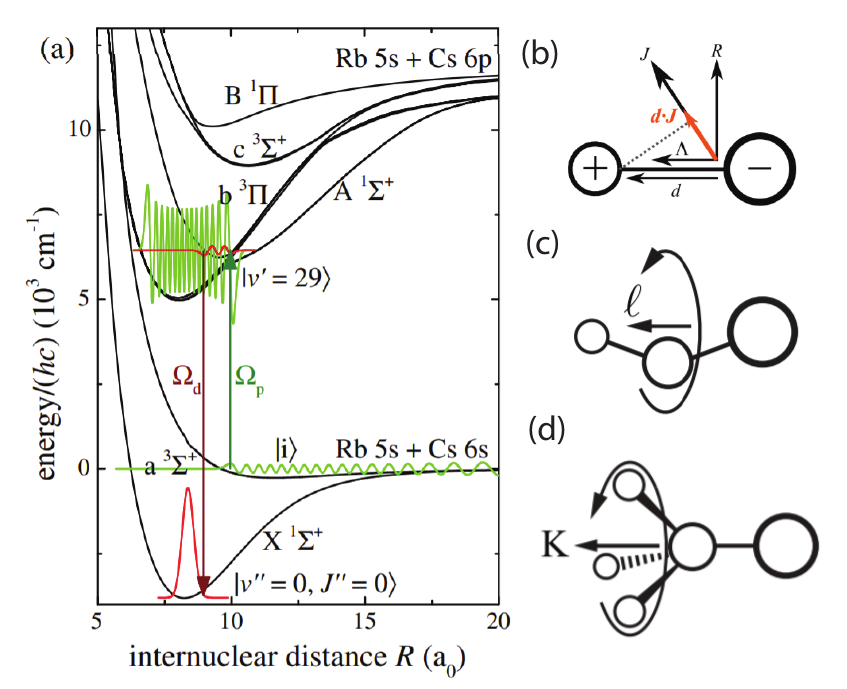}
\caption{\label{fig:structure} (a) Molecular potential energy as a function of interatomic distance for the heteronuclear diatomic molecule RbCs (adapted from \cite{NagerlTakekoshiPRL14_GroundStateRbCs}). The ground electronic state of this molecule is a singlet state represented as $X^1\Sigma^+$. The green and red curves represent the wave functions corresponding to different vibrational states occupied in the process of {light assisted ground-state} molecule formation. Vertical green and red lines denote the laser frequencies involved in the STIRAP process {(Sec. \ref{Sec:VibCoh})}. Parts (b)-(d) are cartoon representations of three classes of molecules discussed in Sec. \ref{sec:MolStructure}; adapted from \cite{Hutzler2020polyatomic}.  (b) Diatomic polar molecule where mixing of $\Lambda$- or $\Omega$- doublets leads to a finite dipole moment in an external field. (c) Linear triatomic molecules where $l$-type doubling occurs due to the bending vibrational mode. (d) Symmetric top molecule with a $K$-type doubling. {In all three cases, generally speaking, mixing between rotational states will also lead to a finite dipole moment.}}
\end{figure}

In the context of quantum control, molecules are also conveniently classified based on their atomic composition and shape since these attributes dictate the molecular symmetries:
\begin{enumerate}
    \item {Homonuclear diatomic molecules, composed of two identical atoms. The electronic states are spin singlets or triplets because of an even total number of electrons. There is an additional label given by the inversion of the electronic wave function through the center of symmetry. This even or odd symmetry is denoted by the subscript $g$ or $u$,
    {As an example, the ground electronic state is $^1\Sigma_g^+$ for the molecules considered in this Perspective}. These molecules are invariant under rotation by any angle about the internuclear axis, contain an inversion center, and fall under the symmetry group $D_{\infty v}$. They possess a single vibrational mode which is the stretching of the bond. {In addition, nuclear exchange symmetry dictates that either only even (bosonic) or only odd (fermionic) values of $J$ are allowed.}}
    \item{Heteronuclear diatomic molecules, composed of two dissimilar atoms.
    {For the molecules covered in this Perspective, the electronic ground states could be either singlet (e.g., $^1\Sigma^+$ for RbCs, Fig. \ref{fig:structure}(a,b)) or doublet (e.g., $^2\Sigma^+$ for CaF or YbCs)}. These molecules lack inversion symmetry, and belong to the point group $C_{\infty v}$.}
    \item{Linear triatomic molecules, composed of three atoms with a linear geometry in the ground state, e.g., {CaOH and SrOH}. These molecules have identical properties to heteronuclear diatomics such as CaF except they have four vibrational modes ($N_{\mathrm{modes}}=3N_{\mathrm{atoms}}-5$). Their bending vibrational mode gives rise to nearly degenerate opposite-parity $l$-doublets
    that allow for full polarization of the molecule at fairly low electric fields ($\sim$100~V/cm, Fig. \ref{fig:structure}(c)).}
    \item{Symmetric top molecules, which are nonlinear but have rotational symmetry about the internuclear axis (e.g., CaOCH$_3$ which belongs to the $C_{3v}$ symmetry group). These molecules have an additional rotational quantum number (denoted by $K$, Fig. \ref{fig:structure}(d)) which defines the rotational angular momentum about the internuclear axis. The lowest-energy rotating state possesses a nearly degenerate opposite-parity $K$-doublet which allows for full polarization at very low electric fields ($\sim$10~V/cm).}
\end{enumerate}

\section{Quantum control of cold molecules}
\label{sec:QuantumControlMolecules}

\subsection{Cold and ultracold ensembles}
\label{Sec:CoolTrap}

The first step toward achieving full quantum control of a molecule involves taming its motional degrees of freedom. At room temperature, the typical speed of a modestly sized molecule in free space is $\sim300$ m$/$s. Static trapping is prohibitive at such high velocities, and spectroscopy is Doppler broadened which complicates manipulation of the {internal} states. The overarching goal is thus to prepare samples of translationally cold molecules such that they can be held in conservative trapping potentials that enable long interrogation times. To attain this level of control, several approaches have been developed.

\subsubsection{\textbf{Molecule assembly and sympathetic cooling}}
\label{Sec:AssemblyOfUltracoldMolecules}

\begin{figure*}[ht!]
\centering
\includegraphics[scale=2]{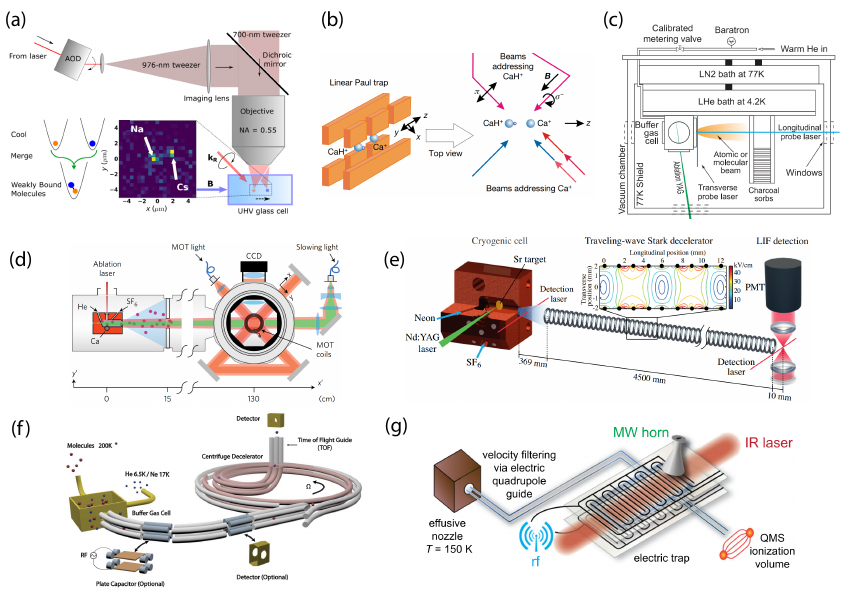}
\caption{\label{fig:Routes} A multipronged approach {to creating} cold and ultracold ensembles of molecules. (a) Atom-by-atom assembly of a single NaCs molecule in the motional ground state of an optical tweezer starting with ultracold atoms. By parallelizing the production in multiple tweezers, arrays with arbitrary geometries can be synthesized. Adapted from \cite{NiLiuPRX19_MolecularAssembly}. (b) A single CaH$^+$ molecular ion cotrapped with a Ca$^+$ ion. The mutual Coulomb repulsion sympathetically cools the molecule and simultaneously enables quantum-logic spectroscopy. Adapted from \cite{LeibfriedChouNature17_MolecularIonCoherentManipulation}. (c) A cryogenic buffer-gas beam (CBGB) source provides a cold flux of molecules created from laser ablation, typically serving as the starting point for direct cooling with external fields. Adapted from \cite{DoyleMaxwellPRL05}. (d) CaF molecules are laser cooled below the Doppler limit and held in a magneto-optical trap. Adapted from \cite{TarbuttTruppeNaturePhys17_CaFBelowDopplerLimit}. (e) A traveling-wave Stark decelerator for SrF. In order to address a wider class of molecules, including polyatomic species, nonradiative methods of cooling and slowing are being explored. Adapted from \cite{HoekstraAggarwal21_SrFDeceleration}. {(f) A cryofuge decelerator for polar molecules. A rotating quadrupole electric guide is used to slow a beam emanating from a {CBGB}. Adapted from \cite{RempeWuScience17_CryofugeColdCollisions}.} (g) Opto-electric Sisyphus cooling of H$_2$CO. A combination of rf {fields} to drive rotational transitions and a strong {electrostatic} trap create an effective potential hill that efficiently removes kinetic energy. Adapted from \cite{ZeppenfeldPrehnPRL16_OptoelectricalCoolingH2CO}.}
\end{figure*}

The tried and true approach involves associating ultracold atoms into molecules using photoassociation \cite{StwalleyPechkisPRA07_85Rb2PAResCoupling,DeMilleSagePRL05_RbCsViaPA,DeMilleBruzewiczNJP14_GroundStateRbCsViaPA,ElliottStevensonPRA16_LiRbfeshbachPA,InouyeAikawaPRL10_coherent,SchreckStellmerPRL12_Sr2,ZelevinskyReinaudiPRL12_Sr2,ZelevinskyLeungNJP21_STIRAP,JonesRMP06} or magnetic Feshbach resonances \cite{NiZhangPRL20_TweezerMagnetoassociation,OspelkausVogesPRA20_magneticfeshbach,KohlerRMP06,ChinRMP10_FeshbachRes}.  The latter is largely applicable to alkali-metal dimers (e.g., KRb \cite{NiSci08}), while the former can work well for nonmagnetic alkaline-earth-metal dimers (e.g., Sr$_2$ \cite{ZelevinskyReinaudiPRL12_Sr2}).  {Both methods utilize fields, either optical or slowly-varying magnetic, to couple atomic scattering states to bound molecular states, in either excited or ground electronic potentials.}  Since the associated molecules inherit the phase-space density of its atomic constituents, molecular gases created this way hold the record for the highest phase-space densities to date \cite{de2019degenerate,DudaBloch_2021_NaKDegenerateFermi,greiner2003emergence,zwierlein2003observation,jochim2003bose}. Three decades of progress in laser cooling and trapping of atoms have now enabled single diatomic molecules to be assembled atom-by-atom in the ground motional state of an optical tweezer (Fig.~\ref{fig:Routes}(a)). Thanks to the tight radial confinement of the tweezer, the motion of individual atoms is quantized and cleanly resolved in frequency, enabling coherent transfer to a bound molecular state \cite{NiYuPRX21_coherent,ZhanHeScience20_coherently}.

Sympathetic cooling of molecules by laser-cooled atoms or ions is a related technique. For neutral particles, the key figure of merit is the ratio of elastic to inelastic collisional cross sections such that cross-species thermalization is effective. Notably, collisional cooling has been recently demonstrated {for} fermionic NaLi molecules with Na atoms where the molecular phase-space density was increased by a factor of 20 \cite{KetterleSonNature20_NaLiNacollisional}, and rapid thermalization has been observed {for} fermionic KRb Feshbach molecules with K and Rb atoms \cite{YeTobiasPRL20_thermalization}. Similar ideas are being explored for CaF \cite{TabuttJurgilasPRL21_collisions,TarbuttLimPRL15_modeling,TarbuttWrightPRR19_microwavetrap}, SrF \cite{DeMilleMccarronPRL18_magnetic,TscherbulMoritaPRA18_Rb_SrF_Cooling}, O$_2$ \cite{NareviciusAkermanPRL17_O2Li}, and OH \cite{HutsonLaraPRL06_Rb_OH_Collisions} with degenerate gases of alkali-metal atoms. For charged particles, the criterion for direct elastic collisions is lifted since this can be effectively mediated through coupled motional modes in a Coulomb crystal. Robust cooling of a molecular ion to its ground motional state can be achieved by applying Doppler and sideband cooling to an atomic ion. This strategy has been shown to be effective, for instance, {in} a linear Paul trap for CaH$^+$ cotrapped with Ca$^+$ (Fig.~\ref{fig:Routes}(b)) \cite{RugangoNJP15_sympatheticCaHion}, MgH$^+$ with Mg$^+$ \cite{SchmidtWanPRA15_efficientMgHion}, and HD$^+$ with Be$^+$ \cite{SchillerAlighanbariNPhys18_HDIonLambDicke}. Moreover, the cotrapped ion can be utilized as an auxiliary particle for quantum logic spectroscopy \cite{WinelandSchmidtScience05} affording high-fidelity, non-destructive molecular state readout \cite{SchmidtWolfNature16_MolecularIonQuantumLogic,LeibfriedChouNature17_MolecularIonCoherentManipulation,WillitschSinhalScience20_quantumnondemo_molecules}. An intriguing middle-ground involves immersing a single molecular ion in a reservoir of ultracold atoms created from a magneto-optical trap (MOT). Such hybrid platforms combine the versatility of ion traps with the ultracold temperatures available in neutrals and are currently being explored \cite{Hudson2016_sympathetic,HudsonPuriScience17_MOTion,Jyothi2016_photodissociation,Jyothi2019hybrid,WillitschDorflerPRA20_rotationalcoolingtheoryN2}.  {While a powerful tool, yielding the lowest molecular phase-space densities and the highest spectroscopic resolution achieved to date, assembly of molecules that relies on ultracold atoms is limited to atomic species that can be laser cooled.  This list is growing, but in practice still largely limited to alkali and alkaline-earth metals.}

\subsubsection{\textbf{Direct laser cooling}}
\label{SubSec:direct_cooling}

Complementary to atom assembly, the phase-space density of a molecular gas can be directly compressed using laser cooling and low-temperature technology. Such a top-down approach has recently been gaining traction, {as it circumvents the limitation of molecular composition based solely on laser-coolable atoms (Sec. \ref{Sec:AssemblyOfUltracoldMolecules})}. Traditional molecular beam experiments typically employ supersonic beam expansion to obtain a source of translationally cold molecules with a narrow velocity spread \cite{Cam2000_supersonicbeamsBook,Aggarwal2021_SrFsupersonic}. This is a straightforward, easily generalizable technique since adiabatic cooling due to the sudden expansion of a gas rests on thermodynamic principles and is therefore agnostic to the finer details of the molecular species. Another well-established technique is that of the cryogenic buffer gas beam (CBGB) source (Fig.~\ref{fig:Routes}(c)) \cite{DoyleHutzlerCR12_BufferGasBeams,DeMilleBarryPCCP11_CryogenicMolecularBeams,HutzlerTakahashiPRR21_CBGBsimulation}. {Compared to supersonic beams, a beam emanating from a CBGB source can be two orders of magnitude higher in brightness while being two times slower \cite{DoyleHutzlerCR12_BufferGasBeams}.} The desired molecular species is typically formed via laser ablation of a solid target followed by a chemical reaction in a cryogenic environment where {the molecules} quickly thermalize with a buffer gas (e.g., helium or neon). Since the removal of kinetic energy only relies on elastic collisions with the buffer gas, the CBGB serves as a versatile low velocity source ($\sim$ 100 ms$^{-1}$) for a wide variety of atoms and molecules. High molecular flux can be achieved in the hydrodynamic enhancement regime, where the molecule has enough time to escape the cell before sticking to the cold cell walls \cite{DoyleHutzler2011_cryogenic,DoyleLu2011_CaHcold}. Furthermore, collisions with the buffer gas allow for cooling the rotational and vibrational degrees of freedom, typically leading to a majority of the molecules occupying the rovibrational ground state. Indeed, this technology is indispensable for magnetic trapping of paramagnetic molecules such as CaH \cite{deCarvalhoEPJD99,DoyleWeinsteinNature98_magnetic}.

\begin{figure}[!]
\includegraphics[width=\textwidth]{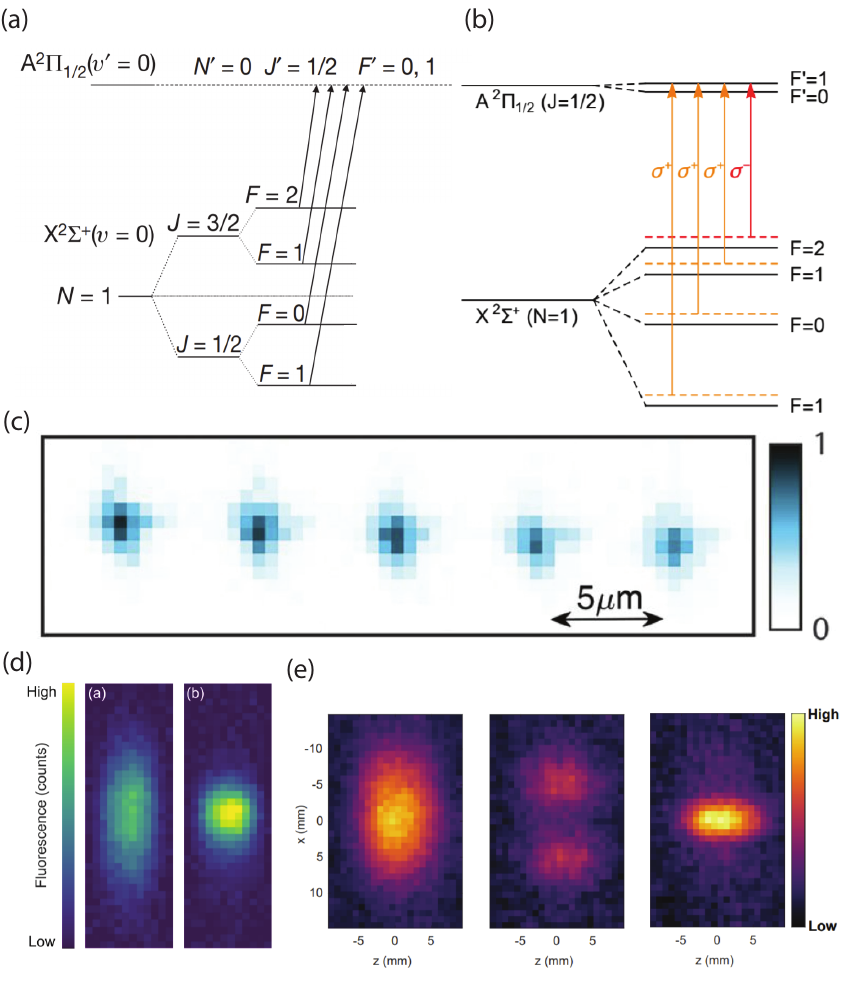}
\caption{\label{fig:laser_cooling} Direct laser cooling of molecules. (a) A typical rotational closure scheme that involves addressing the $N = 1\leftrightarrow0$ {transition} in a $X^2\Sigma^+$ ground manifold (adapted from \cite{DeMilleBarryNature14_SrFMOT}). The two spin-rotation sublevels of the $N=1$ rotational state are denoted as $J=1/2, 3/2$. The hyperfine sublevels within each are denoted by the $F$ {quantum number}. (b) Laser scheme for a dual-frequency DC MOT \cite{TarbuttFitch2021_lasercoolingreview}. The optimized scheme uses {light with} opposite polarizations to address different hyperfine sublevels, ensuring that dark states continue to {participate in optical cycling}. (c) Single CaF molecules trapped in a tweezer array \cite{DoyleAndereggScience19_CaFTweezers}. Molecules are loaded from a DC MOT to an {optical dipole trap} after gray molasses cooling and then transferred to a tweezer array with single occupancy.  (d) 1D MOT compression of a CaOH beam \cite{DoyleBaumPRL20_1DMOTCaOH}. An AC MOT scheme was used to demonstrate magneto-optical forces exerted on a polyatomic molecule.  (e) Sisyphus effect demonstrated for a symmetric top molecule, CaOCH$_3$, {\cite{DoyleMitraScience20_CaOCH3_1DCooling}.}}
\end{figure}

For a long time it was believed that laser cooling of molecules was unfeasible. Electric dipole transitions between rotational states are governed by selection rules {for total angular momentum,} $\Delta J=0,\pm1$, and total parity, $+\leftrightarrow-$. While rotational closure can be ensured by driving $N=1 \leftrightarrow 0$ transitions and applying magnetic fields to destabilize the dark states within the $N=1$ substructure (spin-rotation and hyperfine sublevels, Fig. \ref{fig:laser_cooling}(a)), vibrational transitions are not restricted by selection rules but are determined by Frank-Condon factors ({FCFs,} the overlap integrals {for pairs of} vibrational wave functions). Thus it would seem that any attempt to scatter photons off a molecule will quickly pump population into states off-resonant from the cooling light, creating a bottleneck in the photon cycling rate. It turns out that a sizable subset of molecules possess diagonal FCFs, where the electronic and vibrational degrees of freedom are strongly decoupled such that vibrational relaxation pathways can be closed with reasonable experimental complexity \cite{DiRosaEPJD04_LaserCoolingMolecules,YeStuhlPRL08_PolarMoleculeMOT}. This feature is enabled by a bond between a metal atom like Ca, Yb, or Al and a strongly electronegative ligand like F, Cl, or O$R$ (where $R$ represents any functional group).  {In such a bond,} the electron cloud is localized around the metal atom. In other words, an electronic excitation marginally changes the bond length in these molecules and photon scattering is localized at the metal atom which plays the role of an optical cycling center.

The cold and slow beams emanating from a CBGB source have forward velocities of $\sim150$ m/s. Although slower than for oven-based sources, these velocities are much higher than capture velocities of typical MOTs {(Fig. \ref{fig:Routes}(d))}. To date, all molecular MOTs are preceded by radiative slowing to achieve velocity classes within the MOT capture velocity. The rotational cycling scheme, $N = 1 \leftrightarrow 0$, leads to a larger number of magnetic sublevels in the ground state compared to the excited state. This inverted structure compared to {typical} atomic {cooling} transitions is known as Type-II. Traditional Zeeman slowing on Type-II transitions is feasible but complicated \cite{Petzold_slower_2018}. White light slowing is an alternative slowing method that is not only simpler but also more universal \cite{Zhu_1991_atomic_slowing,DeMilleBarryPRL12_SrFLaserSlowing,Hemmerling_2016_CaF_slowing}. This technique involves frequency broadening a counter-propagating laser such that it is resonant with all velocity classes. Molecules can continuously scatter photons as they travel and decelerate due to radiative forces. Similar slowing forces can be applied by using a technique called chirped slowing \cite{SauerZhelyazkovaPRA14_CaFLaserCoolingSlowing,TarbuttTruppeNJP17_CaFChirpedLaserSlowing} where the frequency of the counter-propagating laser is dynamically tuned to keep a dominant velocity class on resonance while the molecules decelerate.

A more versatile way to accomplish longitudinal slowing is through the use of Stark and Zeeman decelerators to minimize or eliminate the need for optical cycling. When implemented together with CBGBs, such slowing techniques have proven to be very effective. A remarkable demonstration is a 4.5 m long traveling-wave Stark decelerator (Fig.~\ref{fig:Routes}(e)) where SrF molecules are stopped in the laboratory frame as the beam traverses through a series of ring-shaped electrodes, reaching a density comparable to early MOTs of the same species \cite{HoekstraAggarwal21_SrFDeceleration}. Another example is a Zeeman-Sisyphus decelerator \cite{TarbuttFitch2016_ZeemanSisyphusTheory} where molecules are continuously decelerated by the potential landscape created by static magnetic fields, requiring only a handful of photons to optically pump the molecules between weak- and strong-field seeking Zeeman states. This method was demonstrated recently for a beam of CaOH that was continuously decelerated to speeds below 15 ms$^{-1}$ \cite{DoyleAugenbraun2021_zeeman}. {A general slowing technique known as a cryofuge that can slow a wide range of polar molecules has also been developed \cite{RempeWuScience17_CryofugeColdCollisions}. This technique combines a cryogenic source with a centrifuge decelerator and is based on techniques that are agnostic to internal molecular structure. A quadrupole electric field guide is used to confine the molecular beam and a separate, rotating quadrupole guide decelerates the molecules by forcing them up a centrifugal potential barrier \cite{Chervenkov_Rempe_PRL2014_Centrifuge_Decelerator} (Fig.~\ref{fig:Routes}(f)). Using this technique, a beam of CH$_3$F was decelerated from 165 m/s to below 20 m/s.}

Another {cooling} technique known as opto-electric Sisyphus cooling has been successfully applied to formaldehyde, CH$_2$O, where the kinetic energy of the polar molecule is dissipated at the repulsive walls of a boxlike electric trap (Fig.~\ref{fig:Routes}(g)).  {Microwave-driven} rotational transitions to less Stark-perturbed states followed by optical pumping back to the highly Stark-perturbed state result in $>3\times10^5$ molecules at a final temperature of $\sim400$ $\mu$K \cite{ZeppenfeldPrehnPRL16_OptoelectricalCoolingH2CO}. This technique {is highly promising} for cooling complex molecules where cycling thousands of photons is infeasible.

A slow beam of molecules can be trapped in a three-dimensional (3D) MOT using the same radiative forces in combination with a magnetic field gradient (Fig.~\ref{fig:Routes}(d)). A MOT on a Type-II transition is feasible but more complicated than a standard Type-I {atomic} MOT due to dark states. Optimal MOT performance requires these states be destabilized, for example, by allowing the dark state to precess about a magnetic field \cite{DeMilleBarryNature14_SrFMOT}. However, this limits the achievable scattering rates and MOT phase space density. A clever solution to this problem is the use of a time varying, or AC, MOT \cite{TarbuttTruppeNaturePhys17_CaFBelowDopplerLimit,DoyleAndereggPRL17_CaF_RF_MOT,YeCollopyPRL18_YO_MOT_3D}. In an AC MOT, dark states are efficiently {remixed} by simultaneously changing the polarization of the MOT beams from $\sigma^+$ to $\sigma^-$, {along with} the sign of the {magnetic field} gradient. This switching must occur at roughly the scattering rate ($\sim10^6$~s$^{-1}$), requiring radio-frequency engineering of in-vacuum field coils. An alternative pathway is a dual-frequency DC MOT \cite{Williams_2017_dual_freq_MOT}, where two opposite-polarization laser beams constitute each MOT arm (Fig. \ref{fig:laser_cooling}(b)). {For states that are weak-field seeking, the restoring force comes from the polarization component which is red-detuned, akin to a Type-I MOT. But the high-field seeking states remain dark to this polarization component, scattering more efficiently from the other polarization component which is blue-detuned with respect to the transition.} {Dual-frequency DC MOTs have become the preferred choice over AC MOTs for their technical simplicity, although damping forces in a DC MOT are significantly reduced compared to its AC counterpart \cite{DoyleAndereggPRL17_CaF_RF_MOT}.}     

Phase-space compression {beyond the MOT} is required to ensure efficient loading into a dipole trap. The sub-Doppler laser cooling technique of {magnetically assisted} Sisyphus cooling has {become} a rite of passage for newly cooled molecules, due to the ability to exert large {optical} forces without scattering thousands of photons. Sisyphus cooling has been demonstrated for SrF \cite{DeMilleShumanNature10_SrFLaserCooling} and YbF \cite{HindsLimPRL18_LaserCooledYbF}, the latter for a search of the electron EDM \cite{Fitch_YbF_EDM_2020} {(Sec. \ref{Sec:FundSym})}. A {Sisyphus-like} sub-Doppler technique called gray molasses cooling \cite{DoyleCheukPRL18_LambdaCoolingCaF,YeDingPRX20_YOSubDoppler,TarbuttCaldwell_PRL2019_deep_cooling_CaF} has enabled loading {molecules} into optical dipole traps \cite{DoyleAndereggNPhys18_CoolingOpticallyTrappedCaF,DemilleLanginPRL21_LambdaODTSrF}. The $\Lambda$-enhanced gray molasses cooling scheme relies on the existence of a velocity-dependent dark state in a three-level {$\Lambda$} system. Gray molasses cooling works efficiently even in the presence of Stark shifts due to high-intensity trap light, which enables further cooling in dipole traps and loading into optical lattices \cite{YeWu2021_YOlattice} and tweezer arrays \cite{DoyleAndereggScience19_CaFTweezers} (Fig. \ref{fig:laser_cooling}(c)). {As a benefit}, photons scattered during the cooling process can be collected using a high-numerical-aperture objective to enable single-molecule resolution imaging. Finally, further cooling to the ground motional trap state is possible with direct sideband cooling \cite{TarbuttCaldwellPRR20_sideband}. 

The rapid progress in manipulating diatomic molecules gave renewed vigor to experiments extending these ideas to polyatomic molecules. Linear triatomics in {the same $C_{\infty v}$ symmetry group} were the first ones to be explored, {since} rotational closure can be ensured for $N = 1 \leftrightarrow 0$ transitions. A clever choice of species can reduce branching to the three additional vibrational modes. Laser cooling was demonstrated for SrOH \cite{DoyleKozyryevPRL17_SrOHSisyphusCooling} and YbOH \cite{DoyleAugenbraunNJP20_YbOH}, while for CaOH, magneto-optical compression in one dimension \cite{DoyleBaumPRL20_1DMOTCaOH} (Fig. \ref{fig:laser_cooling}(d)) and trapping in a 3D MOT {along with} sub-Doppler cooling was demonstrated \cite{Doyle_Vilas_arXiv_2021_3D_MOT_CaOH}. Linear triatomic molecules {feature} a bending vibrational mode. This long-lived mode can be decomposed into two opposite-parity states, one where the central atom rotates clockwise (CW, angular momentum $+\vec{l}$) on a plane perpendicular to the internuclear axis and the other where the rotation is counter-clockwise (CCW, angular momentum $-\vec{l}$) (Fig. \ref{fig:structure}(c)). These nearly degenerate opposite-parity $l$-doublets can mix and allow the molecule to be fully polarized at electric fields of $\sim100$~V/cm, {which is an advantage for precision measurements for two reasons. First, lower electric fields are simpler to generate. Second, this property affords a better control of
systematic effects, since the fully polarized molecules
present ideal internal co-magnetometers \cite{HutzlerKozyryevPRL17_PolyatomicsForTViolation}.}   

Symmetric top molecules (STMs) offer even longer-lived doublet states. STMs have rigid-body rotation which introduces opposite-parity {energy level} doublets (CW and CCW rotation) known as $K$-doubling (Fig. \ref{fig:structure}(d)). This doublet allows for full polarization of the molecule at fields as low as 1~V/cm. However, the same doubling can result in unwanted loss channels during photon cycling. As an example, the $N=1\leftrightarrow0$ scheme {no longer} guarantees rotational closure, requiring additional rotational repumps. Non-rotating states, however, do not possess $K$-doublets and behave similarly to diatomic and linear triatomic species. These ideas were put into practice with CaOCH$_3$, which is the first symmetric-top molecule to be laser cooled \cite{DoyleMitraScience20_CaOCH3_1DCooling} (Fig. \ref{fig:laser_cooling}(e)).   

The {described} cooling schemes rely on high molecular symmetries to minimize losses to unwanted states. {Generalizing} the techniques to molecules that are devoid of symmetries is of significance for future studies. The simplest of this class of molecules are asymmetric top molecules \cite{KotochigovaKlosPRR20_PolyatomicLaserCooling,KozyryevAugenbraunPRX20_MolAsymmCycling}. Molecules like CaSH and CaOCH$_2$CH$_3$ are bent in their ground state and belong to the $C_s$ symmetry group. This class of molecules can possess three distinct dipole moments that could be independently controlled. The opposite-parity doublets in this case are also $K$-type as for STMs, and rotational closure {can be guaranteed by one or two additional rotational repumps per vibronic transition.} Finally, with a clever choice of ligands, highly diagonal FCFs can be achieved.

{Here} we have described the essential features of diatomic and polyatomic molecules {with respect to} laser cooling. The property that makes these molecules desirable for measurements of fundamental physics is the existence of nearly degenerate, long-lived, opposite-parity doublets ($\Omega$-, $l$- and $K$-type) which, when {combined} with {heavy} nuclei, {enable large enhancement of effective internal electric fields (Sec. \ref{Sec:FundSym}).} More detailed discussions of {molecular} laser cooling {can be found in,} e.g., Refs.~\cite{TarbuttFitch2021_lasercoolingreview,McCarronJPB18_LaserCoolingMolecules}. We also note a related tutorial of molecular cooling methods for precision spectroscopy and tests of physical laws in Ref.~\cite{wall2016_preparation}.

\subsubsection{{\textbf{Lamb-Dicke trapping and Doppler-free spectroscopy}}}
\label{Sec:LD}

{For the most demanding applications, a finite ensemble temperature can introduce deleterious Doppler shifts. Fortunately, Doppler and probe recoil shifts can be strongly suppressed by operating in the Lamb-Dicke regime where the spatial confinement is much tighter than the wavelength of the probed transition, such that the entire probed ensemble samples nearly the same phase of the electromagnetic wave.} Lamb-Dicke spectroscopy \cite{DickePR53_CollisionalNarrowing} is a powerful technique developed for trapped atomic ions \cite{LeibfriedRMP03}, and the principle can be extended to a wide range of neutral and ionic molecular species. While Lamb-Dicke spectroscopy has been routine for atomic clocks \cite{SchmidtLudlowRMP15_OpticalAtomicClocks}, its implementation for molecules became possible only recently. 

For neutral species, spectroscopy in the Lamb-Dicke and resolved sideband regime has been demonstrated in ultracold Sr$_2$ molecules trapped in an optical lattice {(a trapping potential from a standing wave of laser light)}. By probing along the axis of tight confinement, the Lamb-Dicke condition is satisfied for optical transition frequencies of $\sim$500 THz \cite{ZelevinskyMcDonaldPRL15_Sr2LatticeThermometry}, and even more so for Raman rovibrational transitions with frequencies of $\sim$100 MHz to 30 THz driven by co-propagating probe lasers \cite{ZelevinskyKondovNPhys19_MolecularClock,ZelevinskyLeungPRL20_MClockRabi100ms,ZelevinskyLeungNJP21_STIRAP}. While this is a reliable method to achieve Doppler-free spectroscopy, the technical requirements are fairly stringent. High densities can limit the lifetime of the molecular sample due to two-body reactions and the trapping wavelength should be chosen judiciously to avoid accidental one- or two-photon transitions to excited {rovibronic} states. Moreover, ultracold temperatures are typically needed to load an optical trap, which can limit the choice of molecular species. 

For molecular ions, Lamb-Dicke spectroscopy can be realized for a broader set of configurations with comparably relaxed technical requirements. The common strategy is to arrange the ions into narrow strings and probe along the perpendicular (i.e., tight-confinement) direction with a relatively long transition wavelength. This has been achieved for frequency-comb driven Raman rotational transitions in single CaH$^+$ ions \cite{LeibrandtChouScience20}, direct rotational transitions in HD$^+$ ion ensembles \cite{SchillerAlighanbariNature20_HDIonLambDickeQEDTests}, direct mid-infrared vibrational spectroscopy of HD$^+$ ensembles \cite{SchillerKortunovNPhys21_HDIonLambDickeMassRatio,SchillerAlighanbariNPhys18_HDIonLambDicke}, and two-photon vibrational transitions in HD$^+$ ensembles with counter-propagating {(along the trap axis)} probes of closely matched wavelengths \cite{PatraKoelemeijScience20_HDIonProtonElectronRatioPPT}.

\subsection{Molecular vibrations}
\label{Sec:VibCoh}

\begin{figure}[!]
\includegraphics[width=\textwidth]{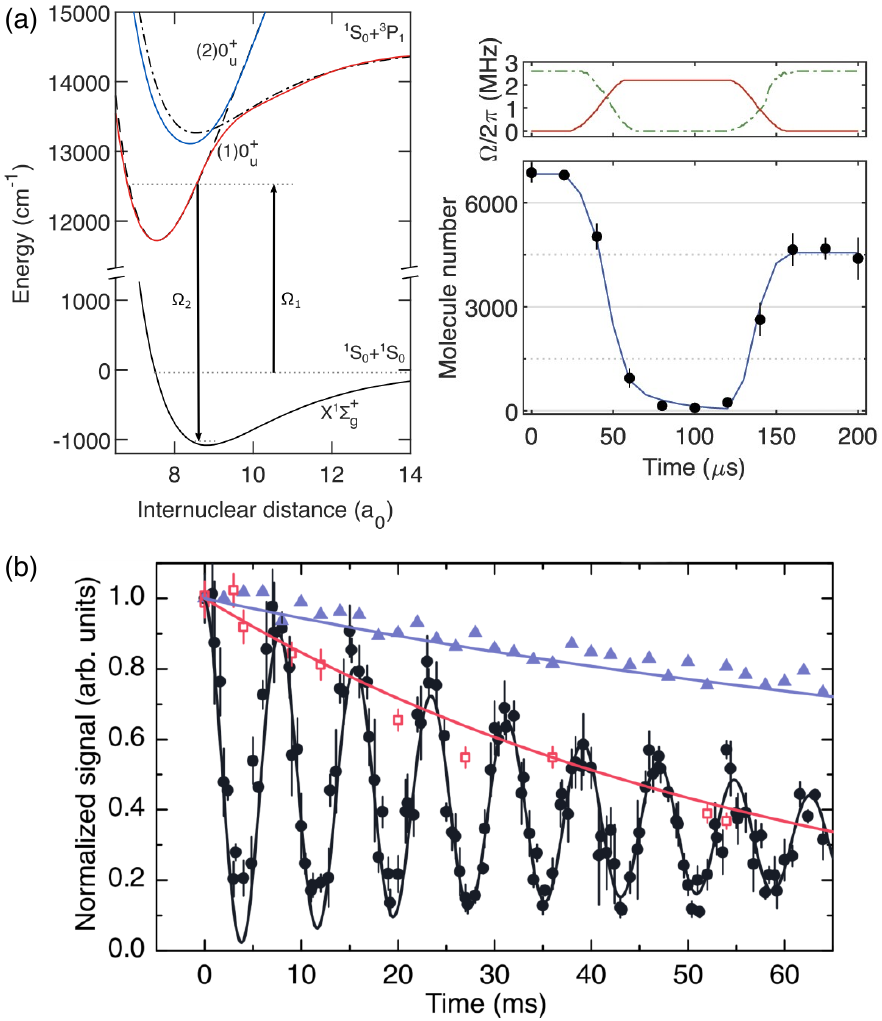}
\caption{\label{fig:vibcontrol} Coherent control of molecular vibrations in $\mathrm{Sr}_2$. The molecules are confined in a {1D} optical lattice and interrogated in the Lamb-Dicke and resolved-sideband regimes. (a) Stimulated Raman adiabatic passage (STIRAP) coherent population transfer to the absolute rovibrational ground state starting from {weakly bound} photoassociated molecules. In these experiments, transfer times are typically $<20\,\mu\mathrm{s}$. To detect the transferred molecules, the STIRAP sequence is reversed as shown. Adapted from \cite{ZelevinskyLeungNJP21_STIRAP}. (b) Rabi oscillations across nearly the entirety of the $X^1\Sigma_g^+$ ground potential achieved by tuning the optical lattice to a magic wavelength for the vibrational clock states. {The upper (blue) and lower (red) decaying traces show the collisional and lattice scattering limitations on the coherence time, respectively.}  Adapted from \cite{ZelevinskyLeungPRL20_MClockRabi100ms}.}
\end{figure}

A common feature of atom-assembled molecules is that they are formed in high-lying vibrational states near their respective dissociation thresholds. In certain cases, the vibrational states with favorable properties for fundamental physics applications {are more deeply bound} (Sec. \ref{sec:temporalvar}, {\ref{sec:NewForces}}). {As an example, intermediately bound vibrational states generally have the largest sensitivity to the electron-to-proton mass ratio variations ($\dot{\eta}$) in a given electronic potential, since the sensitivity is proportional to $(v+1/2)/\rho_v$, where $v$ is the vibrational quantum number and $\rho_v$ is the density of states around state $v$ \cite{rios2020ultracold}. Intuitively, a change in $\eta$ leads to a shift in the fundamental vibrational frequency ($\omega_e$) of the molecule, which becomes amplified for higher-lying vibrational states since the energies grow as $\sim n\omega_e$. However, near the atomic dissociation threshold, the anharmonicity and gentler slope of the potential lead to an increased density of states resulting in a diminished sensitivity.} Alternatively, the sensitivity to $\dot{\eta}$ may be effectively enhanced by probing transitions between accidental near-degeneracies in the vibrational ladders of two different potentials \cite{InouyeKobayashiNComm19_MuDotKRb} (see also Fig.~\ref{fig:qedmudot}(b)). In either case, the weakly bound molecules will have to be de-excited to lower vibrational states.

When transferring population across a large energy difference, coherent methods compare favorably to optical pumping schemes (apart from special cases with large Frank-Condon factors) as the former enable high transfer efficiencies and preserve phase-space density. A powerful and widely employed technique is that of stimulated Raman adiabatic passage (STIRAP). In its simplest implementation, STIRAP utilizes two laser fields, each addressing the same intermediate state from the initial and target states (Fig. \ref{fig:structure}(a)). State-selective transfer is accomplished by controlling the STIRAP laser polarizations. Keeping the lasers on two-photon resonance, the laser intensities are varied sequentially so as to adiabatically maintain the molecule population in a dark state created by the coherent superposition of the laser fields. This serves to minimize scattering losses due to the intermediate state. STIRAP requires the relative coherence time of the laser fields to be longer than the transfer time, which in turn must be longer than the time-scale of the resonantly-driven dynamics set by the respective Rabi frequencies in order to maintain adiabaticity. Fortunately, laser technology in the visible and near-infrared has become so reliable that STIRAP is routine for many experiments working with bi-alkali metal molecules, and recently also for bi-alkaline-earth metal molecules (Fig. \ref{fig:vibcontrol}(a)). STIRAP has also been utilized to efficiently prepare ThO molecules in electron-EDM-sensitive spin states that are aligned to an external electric field \cite{ACMEPandaPRA16_STIRAP_ThO}. It has been suggested as a possible dissociation route toward exotic ultracold gases (e.g., atomic hydrogen) the spectra of which are of interest to the community (e.g., determination of the Rydberg constant and the proton radius) \cite{LanePRA15_HFromBaH}. Typical strategies for maintaining the relative coherence of the STIRAP lasers involve stabilizing both lasers to the same high-finesse optical cavity or low-noise frequency comb such that frequency perturbations are common-mode and suppressed. 

High-precision vibrational spectroscopy of molecules through laser excitation has been demonstrated in a wide range of configurations. Heteronuclear molecules possess a non-zero permanent dipole moment that {varies} with internuclear separation and thus are both microwave and infrared active. The simplest heteronuclear molecule is HD$^+$, consisting of two isotopes of hydrogen and a single electron. Laser excitation of the fundamental vibrational overtone transition ($v=0 \rightarrow 1$) in the ground potential of HD$^+$ was observed {in} the 1970s by Lamb who conjectured that rovibrational spectra of simple molecules can serve as tests of molecular quantum electrodynamics \cite{WingLambPRL76_HDfundamental} (Sec.~\ref{sec:QEDtests}). Recent work with sympathetically cooled HD$^+$ ions fully resolved the hyperfine spectra of this fundamental overtone transition, achieving a fractional uncertainty at the $10^{-9}$ level \cite{SchillerBresselPRL12_HDIonMetrology}. This outstanding precision was later surpassed via an extension of the ``trapped ion cluster transverse excitation spectroscopy'' technique, achieving sub-kHz resolution and a fractional uncertainty of $3\times 10^{-12}$ \cite{SchillerKortunovNPhys21_HDIonLambDickeMassRatio}.

Homonuclear diatomic molecules in their ground potential, on the other hand, are microwave and infrared inactive since a finite dipole moment cannot arise (i.e., the electron density remains evenly distributed around the nuclei at all bond lengths). {Nevertheless}, in both homonuclear and heteronuclear dimers, higher-order coupling between vibrational states can occur through a quadrupole transition or a two-photon process (see also Sec. \ref{sec:forbid}). STIRAP is an example of such a two-photon process and was used to determine the ground potential depth of the RbCs molecule with $4\times 10^{-10}$ fractional uncertainty \cite{CornishMolonyPRA16_RbCsmeasurement}. Here, the leading systematics arise from incomplete knowledge of the dense state structure near the dissociation threshold and uncertainty in zeroing the magnetic field. Both limitations can be circumvented in homonuclear molecules such as Sr$_2$, Yb$_2$, O$_2^+$, or N$_2^+$. Moreover, homonuclear molecules couple very weakly to {room-temperature} blackbody radiation owing to their infrared inactivity, and it is possible to form combinations with zero total nuclear spin that greatly simplify analysis of the spectra. Magic wavelength optical trapping of neutral molecules, which eliminates first-order differential lightshifts from the trap laser, has been demonstrated for Sr$_2$ by tuning the wavelength of the optical lattice near narrow $X^1\Sigma_g^+ \rightarrow a^3\Sigma_u^+$ resonances \cite{ZelevinskyKondovNPhys19_MolecularClock,ZelevinskyLeungPRL20_MClockRabi100ms}. This enabled the observation of vibrational transitions with quality factors {of} $\sim10^{12}$ and two-photon Rabi oscillations persisting for $\sim100$ ms between vibrational states spanning the depth of the ground potential (Fig~\ref{fig:vibcontrol}(b)). A similar idea was proposed for the fundamental vibrational overtone excitation in YbLi where far off-resonant magic wavelengths are predicted \cite{KajitaPRA11_YbLiMagicTrappingProposal}. The ability to coherently manipulate vibrational superpositions opens up the possibility of engineering highly precise vibrational molecular clocks (see also Sec. \ref{sec:forbid}).
 
\subsection{Molecular rotations \label{Sec:RotControl}}

\begin{figure}[!]
\includegraphics[width=\textwidth]{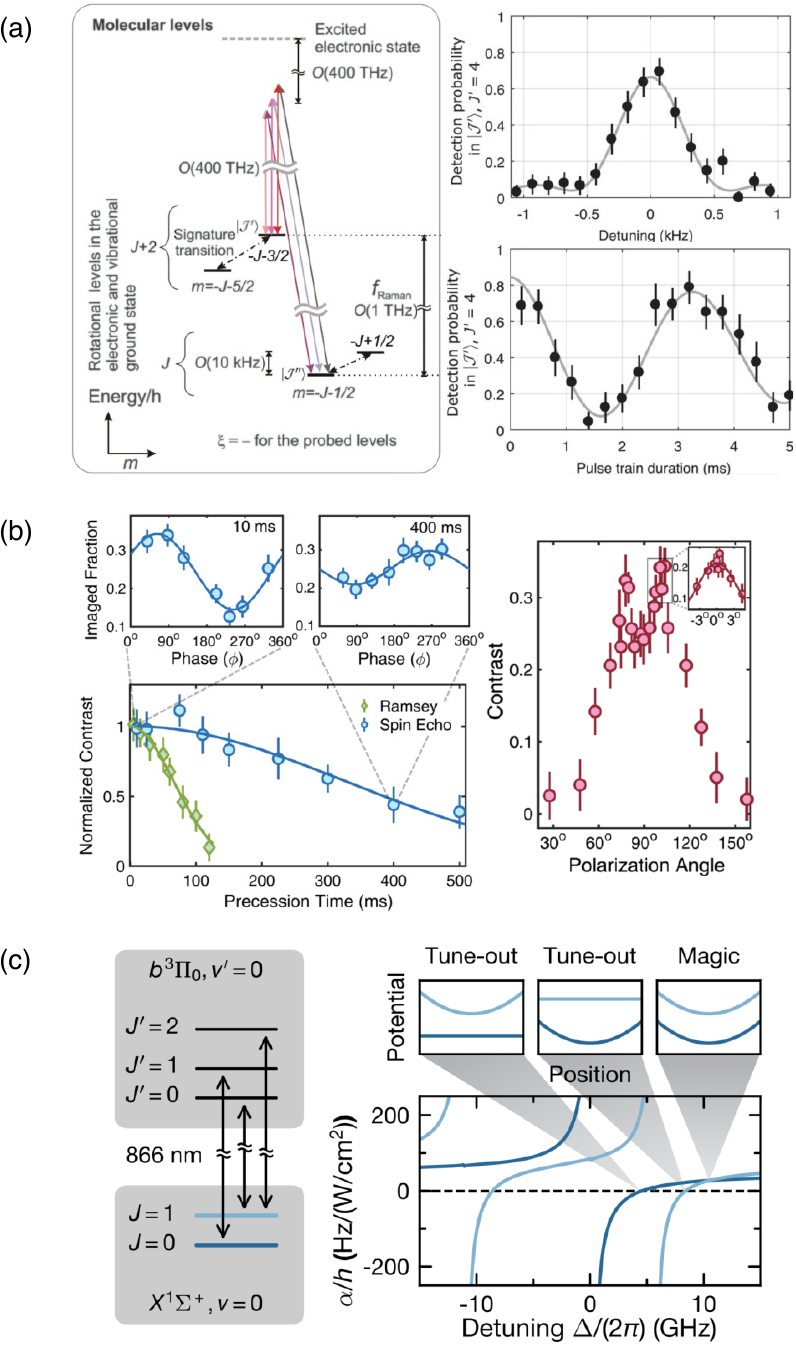}
\caption{\label{fig:rotcontrol} Coherent control of molecular rotations. (a) High-resolution THz-scale rotational spectroscopy and Rabi oscillations in CaH$^+$. The Raman transition is driven with an optical frequency comb. Adapted from \cite{LeibrandtChouScience20}. (b) Half-second scale rotational coherence measured for a single CaF molecule in an optical tweezer, {obtained} by tuning the polarization of the tweezer to special values with respect to an applied magnetic field. Adapted from \cite{DoyleBurcheskyPRL21_rotationalcoh100ms_CaF}. (c) A scheme for engineering magic and tune-out wavelengths for optically trapped bi-alkali metal molecules using a vibronically excited state with a narrow linewidth, shown for NaK. Adapted from \cite{BausePRL20_MagicTuneOutNaK}.}
\end{figure}

The rotational spectra of molecules are accessible with electromagnetic radiation typically in the microwave range ($\sim$10 GHz). While the focus of this Perspective is on tests of fundamental laws, controlling a molecule's rotational degree of freedom is also pertinent for a number of applications such as quantum information and state-controlled chemistry \cite{KochRMP19}. As such, the techniques for rotational control are {quite} well developed.

In Sec. \ref{Sec:CoolTrap} we discussed how the kinetic energy of a molecular ion can be {reduced} through sympathetic cooling with a cotrapped, laser-cooled atomic ion. In heteronuclear species, however, strong coupling to the room-temperature blackbody radiation leads to a thermal redistribution for the rotational {populations}. To overcome this, schemes have been devised to prepare rovibrationally pure samples of molecular ions. One successful approach involves optically pumping $N\geq 1$ rotational levels to $N=0$ with lasers, thanks to favorable Frank-Condon factors. Experimental demonstrations of such rotational cooling were reported for hydrides (e.g., HD$^+$ \cite{SchillerSchneiderNatPhys10_RotcoolingHD}, MgH$^+$ \cite{DrewsenStaanumNPhys10_MgHIonRotationalLaserCooling}, and AlH$^+$ \cite{OdomLienNatComm14_RotcoolingAlHbroadband}) and oxides (e.g., SiO$^+$ \cite{OdomStollenwerkPRL20_SiOIonRotationalCooling}), reaching effective internal state temperatures of $\sim$1--10 K. Non-optical approaches, such as that explored for HD$^+$ using a helium buffer gas  \cite{DrewsenHansenNature14_MgHIonRotationalCoolingHe}, {are} capable of achieving similar temperatures and may prove to be more favorable than optical pumping for infrared-inactive species such as N$_2^+$ \cite{WillitschDorflerPRA20_rotationalcoolingtheoryN2}. Besides cooling, repeated cycles of optical pumping can be used to prepare molecular states with large rotational quantum numbers called ``super rotors''. This was elegantly demonstrated in SiO$^+$ where super rotors with narrow distributions around $N\sim100$ were engineered and used to probe a previously unobserved molecular potential \cite{OdomAntonovNComm21_preciselyspunrotors_SiO}.

Rotational cooling may be sidestepped by prepending the spectroscopy with a quantum logic sequence that projects the molecule in a pure rotational state. At their heart, quantum logic protocols rely on state-dependent forces to act as an information bus between multiple particles. To list a few concrete examples, these forces could be mediated through optically induced dipoles \cite{SchmidtWolfNature16_MolecularIonQuantumLogic}, shared motional modes \cite{LeibfriedChouNature17_MolecularIonCoherentManipulation}, or a combination of both \cite{WillitschSinhalScience20_quantumnondemo_molecules}. The latter is simple to understand and its salient features are illustrated as follows. A blue-sideband optical transition is first driven in the molecule, the motional modes of which are coupled to that of a cotrapped atom. If successful, the atom (originally in its motional ground state and an excited dark electronic state) gains an additional {motional} quantum, and a de-excitation to a bright electronic state is now {energetically} allowed. Since rotational transition frequencies are unique and well resolved, subsequent detection of fluorescence from the atom heralds the state of the molecule. In this manner, pure rotational states in CaH$^+$ were coherently manipulated using a frequency comb to directly drive THz transitions (Fig.~\ref{fig:rotcontrol}(a)) \cite{LeibrandtChouScience20}. Extension of these ideas allowed for the observation of quantum entanglement between a molecular ion and an atomic ion with a fidelity of $\sim80\%$ \cite{LeibrandtChouScience20}.

{Long rotational coherence times have been recently observed in laser-cooled CaF molecules.} Initial experiments conducted {with} bulk gases of CaF in free space or in magnetic traps demonstrated coherence times of a few milliseconds by working with rotational states having similar magnetic moments \cite{TarbuttWilliamsPRL18_BTrappedLaserCooledCaF,TarbuttCaldwellPRL20_longrotationalcoherence,CornishBlackmoreQST19_RotationalCoherence_CaF_RbCs}. Another group approached this differently by loading single CaF molecules into an array of optical tweezers \cite{DoyleBurcheskyPRL21_rotationalcoh100ms_CaF}. Here, mixed quantization arising from the competition between the applied magnetic field and the electric field polarization of the tweezer leads to a crossover behavior in the lightshift experienced by the molecule. The differential lightshift initially increases linearly with tweezer power, but at higher powers the hyperpolarizability of one of the rotational states begins to dominate which changes the sign of the trend, implying the existence of a turning point. By tuning the polarization angle of the tweezer light (Fig.~\ref{fig:rotcontrol}(b)) and operating at applied magnetic fields and tweezer intensities where the differential lightshift is insensitive to fluctuations in tweezer power, light-molecule rotational coherence times of $\sim100$ ms {were achieved}, and up to {$\sim0.5$ s} with a spin-echo pulse. 

For polyatomic molecules, magic-field rotational spectroscopy was recently demonstrated with CH$_2$O in an electrostatic trap \cite{ZeppenfeldPrehnPRL21_MagicSpectroscopyCH2O}. Here, the key insight lies in the observation that the average alignment of a symmetric top molecule's symmetry axis and an applied electric field is proportional to $KM/J(J+1)$, and is unchanged for $J\rightarrow J+1$, $M\rightarrow M+1$, under the condition $\abs{M}=J/2$. While CH$_2$O is a planar asymmetric top molecule, deviations from the symmetric top structure are negligible {in terms of} the rotational splittings. Competition between $K$-doubling and higher-order Stark shifts leads to a region of electric field strengths where the Stark shifts are first-order insensitive to variations in the electric field. Using opto-electrically cooled molecules {at} 2.6 mK, linewidths as narrow as a few kHz were observed for the transition between the pair of rotational states satisfying these conditions.

More generally, for $N=0\leftrightarrow 1$ {transitions} within the same {vibronic} manifold, the differential lightshift can be minimized by tuning the polarization of the trapping laser to the magic angle of 54.7$^{\circ}$. The origin of this angle can be understood from the rank-2 structure of the polarizability of a particle; i.e., a rank-2 tensor can be decomposed into a symmetric diagonal ``scalar'' term, an anti-symmetric ``vector'' term, and a symmetric traceless ``tensor'' term. Lightshifts, on the other hand, are scalars and thus the contraction of the tensor part of the polarizability with the applied electric field must be proportional to the Legendre polynomial $P_2(\cos\theta) = (3\cos^2\theta -1)/2$, which is zero at a polarization angle of $\theta \approx 54.7^{\circ}$ with respect to the quantization axis for $N\neq 0$. For the scalar state $N=0$, the tensor term is identically zero. This observation was applied to extend the rotational coherence times in ultracold bulk gases of KRb \cite{JinNeyenhuisPRL12_KRbAnisotropicPolarizability} and NaK with a modest applied electric field to decouple the nuclear spins from rotational motion \cite{GohleSeesselbergPRL18_RotationalCoherenceNaK}. Related experiments involving high resolution microwave spectroscopy or coherent Rabi flopping have also been performed in RbCs \cite{CornishBlackmore2020_hyperfinecoherent_RbCs,CornishBlackmoreQST19_RotationalCoherence_CaF_RbCs,JiaGong2021_microwaveRotational_RbCs,JiaJi2020_microwaveROT_RbCs}. Alternatively, instead of controlling polarization and applied static fields, the differential lightshift can be eliminated by engineering magic wavelength optical traps similar to those demonstrated for Sr$_2$ \cite{ZelevinskyKondovNPhys19_MolecularClock,ZelevinskyLeungPRL20_MClockRabi100ms}. In bi-alkali metal molecules, this can be accomplished by tuning the trap wavelength near a narrow $X^1\Sigma^+ \rightarrow b^3\Pi$ transition with large Frank-Condon factor (Fig.~\ref{fig:rotcontrol}(c)). Measurements for NaK \cite{BausePRL20_MagicTuneOutNaK} and NaRb \cite{WangHe21_NaRbMagic}, and calculations for RbCs \cite{CornishGuanPRA21_magic}, all indicate favorable magic trapping conditions where the trap wavelengths can be far-detuned enough such that near-resonant scattering is not detrimental. 

\subsection{Nuclear and electronic spins \label{Sec:HFScontrol}}

\begin{figure*}[!]
\includegraphics[width=0.8\textwidth]{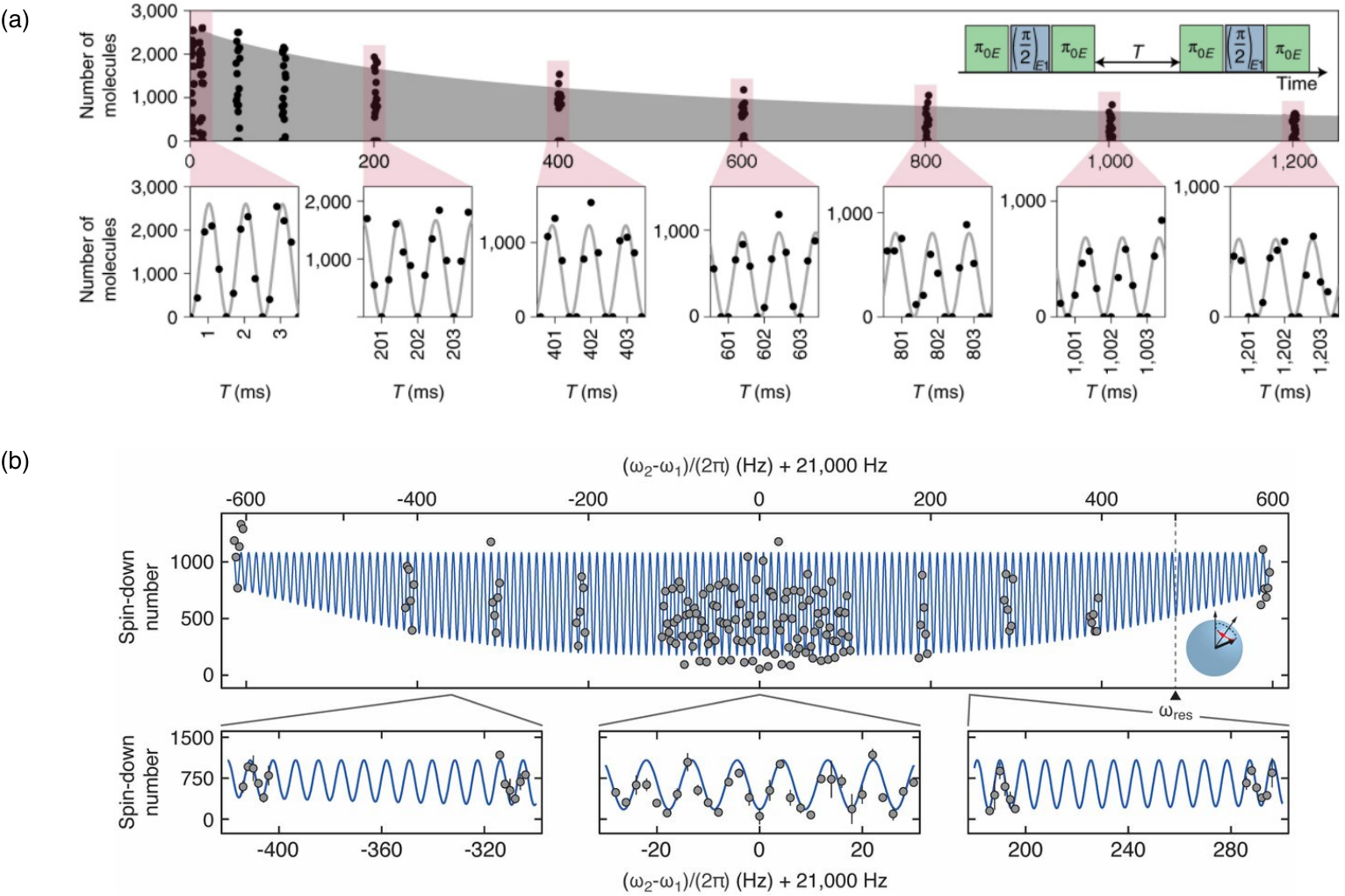}
\caption{\label{fig:hfscontrol} Coherent control of nuclear spin states in molecules. (a) Ramsey spin precession of hyperfine qubits in the rotational ground state of RbCs. Decoherence-free fringes were obtained by tuning the polarization of the optical dipole trap to the magic angle $\sim54.7^\circ$ relative to the quantization axis. Adapted from \cite{CornishGregoryNPHYS21_Robuststorage_RbCs}. (b) Ramsey spectroscopy of the hyperfine structure in NaK, similarly driven via two microwave photons. Second-scale coherence resulted in a Hz-level determination of the hyperfine splitting. Adapted from \cite{ZwierleinParkScience17_NaKNuclearSpinCoherence}.}
\end{figure*}

For a majority of molecular species, the total nuclear spin $I\ne0$. This implies that the hyperfine interaction leads to further substructure within each rotational state.  The same selection rules for transitions between $m_F$ hyperfine sublevels apply for molecules as for atoms, and transitions typically lie in the radio frequency or microwave regimes.

Because hyperfine transitions involve flipping the spin of a nucleon, they are much more robust against external perturbations compared to pure rotational or electronic transitions, and therefore are particularly suited for maintaining quantum superpositions over long periods of time. In fact, second-scale coherence times have been shown to be realistic. Coherent manipulation of hyperfine states and rigorous investigation of the sources of decoherence in ultracold bi-alkali metal molecular species have been done for RbCs \cite{CornishGregoryNPHYS21_Robuststorage_RbCs,CornishGregoryPRA16_controlling_RbCs}, NaK \cite{ZwierleinParkScience17_NaKNuclearSpinCoherence,ZwierleinWillPRL16_coherentmicrowave_NaK} and NaRb \cite{WangLin_NaRbsecondscalecoherence,WangGuoPRA18_high_NaRb}. In these experiments, STIRAP prepares a rovibrationally pure sample in the ground state. Depending on the STIRAP laser polarizations and the initial state, additional microwaves may be needed to transfer population to the hyperfine state of choice. Thereafter, two microwave photons form a three-level {$\Lambda$} system consisting of two ground hyperfine states and an excited rotational state. By slightly detuning the microwaves from two-photon resonance and inserting a dark precession time between the microwave $\pi/2$-pulses, Ramsey fringes were measured to persist at the second scale enabling Hz-level frequency resolution (Figs.~\ref{fig:hfscontrol}(a,b)). The experiment with RbCs measured an {impressive} lower bound for the coherence time of 5.6 s by utilizing a tensor-shift-free magic-angle optical trap (Sec. \ref{Sec:RotControl}).

Molecular ions are in a prime position for observing long coherence times due to the long trapping lifetimes afforded by deep traps and low heating rates. Notably, in an experiment to search for an electron EDM with HfF$^+$ molecules, Ramsey fringes due to electron spin precession between the $m_F = \pm 3/2$ stretched states were measured out to $\sim$ 1 s. Here, the $\pi/2$-pulses were done by ramping down the rotating electric field of the trap for a short duration ($\sim$1 ms) which briefly increases the rotationally-induced coupling between the states \cite{CornellCairncrossPRL17_JILAeEDMI}. Making use of well-resolved {photofragment} angular distributions after state detection via photodissociation, an order of magnitude improvement in statistics for the EDM measurement {was gained}, reaching the quantum projection noise limit \cite{CornellZhouPRL20_MolecularIonOrientedPD}.

\section{Probing fundamental symmetries}
\label{Sec:FundSym}

Fundamental particles are expected to have nearly symmetric charge distributions within the SM. This manifests as nearly vanishing EDMs for electrons, protons, and neutrons \cite{ChuppRMP19_EDMs}.  However, many beyond-SM theories that attempt to explain the observed matter-antimatter asymmetry predict the violation of parity and time-reversal symmetries and hence $CP$ violation, in light of strict $CPT$ symmetry.  This phenomenon can be manifested as a non-zero {permanent} EDM $\vec{d}$ of a particle with a magnetic moment $\vec{\mu}$, since collinear $\vec{d}$ and $\vec{\mu}$ cannot respect both the $P$ and $T$ symmetries.  This new $T$ violation could be several orders of magnitude higher than the SM expectation \cite{CesarottiJHEP19_eEDMInterpretation}.  Moreover, the yet undiscovered $T$-violating particles that give rise to the EDMs can have masses well into the TeV range, potentially beyond the direct reach of experiments such as the Large Hadron Collider.  Experimental searches for new sources of $T$ violation are therefore very well motivated and carry a high discovery potential.  {These studies} 1) directly open a window into the mysteries of the Universe such as matter-antimatter asymmetry; 2) offer a nearly zero-background {detection} platform due to small expected SM values of EDMs; and 3) extend the reach of large accelerators at a much smaller scale and cost. 

To detect tiny EDMs, large {effective} electric fields need to be applied in the direction perpendicular to {the particle's spin}, leading to precession of the spin axis about the field.
While neutron EDMs can be measured directly \cite{AbelPRL20_nEDM}, applying large fields is challenging for bare electrons or protons due to their charge. While in a pure Coulomb system, such as an atom or molecule with {nonrelativistic} point-sized {nuclei and electrons}, an EDM would be undetectable, the relativistic motion of the electrons and the finite nuclear size ensure detectable and, in fact, enhanced contributions to the electron and nucleon EDMs, respectively \cite{SchiffPR63_EDMs,HindsPhysScr97_TSymmetryWithMolecules,FortsonPT03_EDMSearch}.  Therefore, EDM experiments are usually performed with atoms and molecules (neutral or singly-charged).  A nucleon EDM would lead to a nuclear Schiff moment, a dipole-moment-like property which is particularly enhanced in nonspherical nuclei.  Both types of moments are enhanced in heavy nuclei, hence the existing and proposed experiments employ heavy atomic constituents such as Th, Tl, Yb, Ra, {Hg, and Hf}.

\begin{figure}[!]
\includegraphics[width=\textwidth]{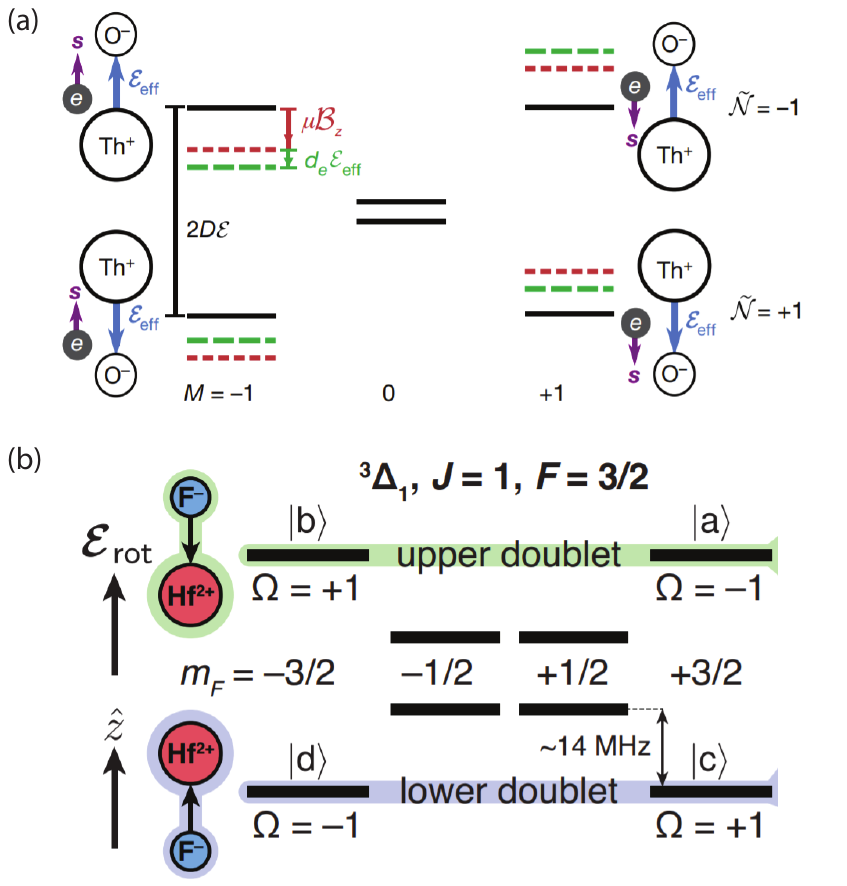}
\caption{\label{fig:EDM} Opposite-parity doublets {employed in} electron EDM (eEDM) searches with molecules. (a) The ACME experiment utilizes the near degenerate $\Omega$-doublets present in the $H$ state of the ThO molecule. The fully polarized molecule can be found in two configurations, where the shift due to the eEDM ($d_e\mathcal{E}_{\mathrm{eff}}$) is either parallel or antiparallel to the Zeeman shift ($\mu\mathcal{B}_z$). Adapted from \cite{ACMENature18_ACMEIIeEDM}. (b) EDM measurement using the molecular ion HfF$^+$. {The interaction between the molecule dipole moment and a rotating electric field $\mathcal{E}_{\mathrm{rot}}$ give rise to Stark-split states.} The Stark doublets have a structure similar to $\Omega-$doublets. Adapted from \cite{CornellCairncrossPRL17_JILAeEDMI}. }
\end{figure}

In recent years, molecule-based experiments have begun surpassing other measurements in providing the most stringent bounds on $T$ violation beyond the SM \cite{YeCairncrossNatRevPhys19_TSymmAtomsMolecules}.  The main key to this success is the ability to {fully} polarize molecules with much lower electric fields than atoms.  If the applied fields can be reduced to the level of some kV$/$cm or below, they are feasible to control in the lab, and give rise to manageable levels of systematic effects.  Using rotational states of polar diatomic molecules typically satisfies this condition.  The success of molecule experiments would not be possible without the newly developed capabilities in controlling the motional and internal quantum states of neutral and ionic molecules (Sec. \ref{sec:QuantumControlMolecules}). The crucial methods involve the use of cycling optical transitions for internal state cooling and high-fidelity fluorescence detection in neutral molecules, as well as state-resolved photodissociation for ionic species. Together with the technology of slow, bright cryogenic molecular beam sources \cite{DoyleHutzlerCR12_BufferGasBeams,DeMilleBarryPCCP11_CryogenicMolecularBeams}, these approaches to quantum control make molecules a highly competitive system to atoms, despite {generally lower numbers of particles available for measurements. Typical CBGB beam brightness is $\sim10^{10}$ molecules$/$pulse over separated pulses that are $\sim10$~ms in duration, while typical oven-based atomic sources provide continuous fluxes of $\sim10^{12}$ atoms/s.  Nonreactive atoms can also be interrogated in vapor cells at relatively high densities.  Nevertheless, motional and state control of molecules is rapidly improving (Sec. \ref{sec:QuantumControlMolecules}).
}

The basic principle of an EDM measurement is the following.  First, the molecular beam is pumped via optical and/or microwave fields to populate a ``science state" where, in the quasiclassical picture, the spin axis is initialized orthogonally to the electric-field ($\vec{\mathcal{E}}$) axis along which the molecules are aligned.  Next, the molecules traverse through a region where they are polarized by $\vec{\mathcal{E}}$.  Their spins precess about the field {axis} at a rate proportional to $\pm d\mathcal{E}_{\mathrm{eff}}$, where $d$ is the $T$-violating moment, $\mathcal{E}_{\mathrm{eff}}$ is the effective intramolecular electric field that can reach several GV$/$cm, {and the sign is based on the choice of the internal state.} Lastly, the population difference in the two science states with opposite {sign of precession} is measured using state-selective optical fluorescence or, for ions, photodissociation.  The difference between these populations provides direct access to $d$ after a careful consideration of systematic uncertainties.   

Paramagnetic molecules with an unpaired electron are most sensitive to the electron EDM, while diamagnetic molecules with nuclear spin can detect nuclear Schiff moments.  Nearby opposite-parity $\Omega$-doublets of diatomic molecules, as in the $H$ state of paramagnetic ThO (Fig. \ref{fig:EDM}(a)), allow the molecules to be fully polarized at fields of $\sim10$ V$/$cm. In this case, the $\sim2$ ms science-state lifetime limited the spin precession time.  Nevertheless, the ACME experiment has provided the most stringent bound on the electron EDM to date, $\abs{d_e}<1.1\times10^{-29}$ $e\cdot$cm  \cite{ACMENature18_ACMEIIeEDM}.  The next generation of this experiment is going to employ the $Q$ state of ThO, with a $>60$ ms lifetime, for greater statistical sensitivity.  This and other improvements promise to gain at least an order of magnitude in sensitivity in the near future \cite{ACME_Panda_2019,Wu_ACME_2020}.

In addition, work is underway to measure the nuclear Schiff moment with diamagnetic $^{205}$TlF molecules.  This measurement is poised to improve the constraints on the proton EDM and the $\theta$ parameter of quantum chromodynamics by one to two orders of magnitude.  Here the science states comprise part of the electronic ground-state hyperfine structure, and therefore the spin precession coherence time is limited by the length of the interaction region and divergence of the molecular beam \cite{KawallGrasdijkQST21_CeNTREX,RamseyWilkening84_PTViolationsTlF,HindsCho91_TlFMeasurement}.  The science states are chosen to suppress and control systematic effects to the highest possible degree:  for example, the directions of the effective intramolecular magnetic and electric fields relative to each other and to the laboratory frame can be reversed simply by detecting different hyperfine components.

Trapped molecular ions, unlike beams of neutral molecules, in principle allow long ($\sim1$ s) interaction times.  The experiment with HfF$^+$ provided a bound on the electron EDM consistent with the ACME experiment \cite{CornellCairncrossPRL17_JILAeEDMI}. Here the ions are directly polarized by small electric fields ($\sim10$ V$/$cm) and the opposite-parity states are Stark doublets (Fig. \ref{fig:EDM}(b)).  However, the constraints of ion trapping prevent loading of large arrays due to repulsive Coulomb forces.  Hence the precision limitation stems from the molecule number rather than interaction time.  Nevertheless, shot-noise-limited detection was achieved with hundreds of ions \cite{CornellZhouPRL20_MolecularIonOrientedPD}.

Much longer interaction times can be achieved for laser-coolable molecules trapped in a 3D optical lattice.  An example of such a molecule with a heavy nucleus is YbF. The techniques of laser cooling detailed in Sec. \ref{SubSec:direct_cooling} could be employed to trap a dense 3D lattice of YbF molecules \cite{Fitch_YbF_EDM_2020}.  The trade-off is that the lack of nearly degenerate opposite-parity states available in this class of molecules leads to larger required electric fields of several kV$/$cm.

Looking into the future, polyatomic molecules could offer the best of both worlds in terms of longer lifetimes and easily polarizable states. Experiments with YbOH and YbOCH$_3$ are expected to achieve full polarization at moderate electric fields \cite{DoyleAugenbraunNJP20_YbOH,Hutzler2020searches}. Some atomic constituents that are highly sensitive to physics beyond the Standard Model, such as $^{225}$Ra, lack the requisite electronic structure to support this property in a diatomic molecule configuration.  As part of a polyatomic molecule, however, this property can be restored \cite{HutzlerKozyryevPRL17_PolyatomicsForTViolation}. Moreover, some molecules provide useful internal comagnetometers for mitigation of systematic effects.  For example, radioactive (half-life of 15 days) polyatomic symmetric-top molecular ions $^{225}$RaOCH$_3^+$ \cite{JayichFanPRL21_RaPolyatomicIons} have been proposed for sensitive Schiff moment searches in a few-particle regime \cite{HutzlerYuPRL21_RaOCH3Ions}.  The octupole deformation of the $^{225}$Ra nucleus and the resulting nuclear state structure \cite{GaffneyNature13_PearShapedNuclei} lead to a significant enhancement of the Schiff moment sensitivity compared to spherical nuclei used in state-of-the-art measurements, such as $^{199}$Hg \cite{HeckelGranerPRL16_HgEDM}.  This paramagnetic molecular ion is relatively insensitive to magnetic field noise but is strongly sensitive to $CP$ violation in the nuclear sector, although the high sensitivity and very long coherence times have to overcompensate for the low particle number. The complexity of polyatomic molecules does not present the same challenge for precision measurements in ion traps \cite{CornellCairncrossPRL17_JILAeEDMI,CornellZhouPRL20_MolecularIonOrientedPD} as it does with laser-cooled neutral molecules \cite{DoyleMitraScience20_CaOCH3_1DCooling}, since optical cycling is not required.

\section{Probing QED and possible new forces}
\label{Sec:QEDNewForces}

As the precision of molecular spectroscopy has reached the part-per-trillion (ppt) level, it is now possible to use diatomic molecules as benchmarks for QED calculations, and as quantum mechanical test systems for new physical forces and interactions.

\subsection{QED tests and fundamental constants\label{sec:QEDtests}}

The smallest molecules can serve as excellent benchmark systems for high-precision tests of QED.  Conversely, when combined with state-of-the-art QED, they can yield precise values of fundamental parameters, and may offer a glimpse of possible new physics \cite{UbachsSalumbidesPRD13_FifthForcesMolecPrecisMeasts,UbachsSalumbidesNJP15_ExtraDimensionsMolecSpectrosc}.

Hydrogen molecules are, of course, the simplest diatomics.  Their energy levels can be fully described using the methods and basic parameters of QED $-$ the Rydberg and the fine structure constants $-$ as well as the electron-to-proton mass ratio, $\eta=m_e/m_p$.  Effects of the weak force such as parity violation have not yet been observed in molecular systems. Femtometer-scale strong interactions are important mainly to the extent that they set the strengths of nuclear magnetic moments and hyperfine structure, as well as the nuclear charge radius, while gravity is too weak to measurably influence molecular spectra.

Mixed isotopologues such as HD offer weak electric-dipole vibrational transitions, versus the purely quadrupole transitions in H$_2$, D$_2$ and T$_2$ \cite{SalumbidesTrivikramPRL18_T2QED}, thus extending the opportunities for high-precision spectroscopy.  The ionic HD$^+$ molecule is a one-electron, three-body system, and thus even more theoretically tractable via $ab~initio$ calculations than the neutrals.

Rovibrational spectroscopy of hydrogen molecules \cite{UbachsDickensonPRL13_H2VibrPrecisMeast,UbachsCozijnPRL18_HDVibrationalMetrology,HuTaoPRL18_MassRatioHDLambDip}, combined with QED calculations, can yield the mass ratio $\eta$.
A variety of methods used to obtain fundamental constants is important to address measurement discrepancies, such as one that arose recently in measurements of the proton mass with cotrapped protons and highly-charged carbon ions \cite{SturmHeissePRL017_ProtonMass}.
Recent spectroscopy of HD yields an experimental precision of $10^{-9}$, pending some necessary improvements in theoretical accuracy \cite{HuTaoPRL18_MassRatioHDLambDip}.  This precision is approaching the CODATA value within roughly an order of magnitude \cite{TaylorTiesingaRMP21_CODATA18}.

The dissociation energy of molecular hydrogen is a fundamental quantity suitable for benchmarking QED calculations \cite{MerktLiuJCP09_H2IonizDissoc,JungenChengPRL18_H2DissocSubMHz}.  To compute the binding energy with an inaccuracy well below $10^{-9}$, it is necessary to consider the Born-Oppenheimer energies along with adiabatic, leading nonadiabatic, relativistic, and QED corrections, which have been subject of theoretical efforts over the past decade \cite{PachuckiPRA10_H2BO,KomasaPachuckiJCP14_H2Adiabatic,KomasaPachuckiJCP15_H2Nonadiabatic,KomasaPachuckiJCP16_H2SchrodingerEq,PachuckiPuchalskiPRL16_H2GroundStateAlpha6Corrections,PachuckiPuchalskiPRL19_H2NonadiabaticQED}.  Recent efforts resulted in some discrepancies between experiments and theory \cite{PachuckiPuchalskiPRA17_H2RelativisticDiscrepancy}, highlighting the need for continuing improvements in precision on both fronts.  The hydrogen dissociation energy has been indirectly measured to a precision of 750 kHz (or $7\times10^{-10}$) using a combination of three energies \cite{JungenChengPRL18_H2DissocSubMHz}.

Molecular spectroscopy of hydrogen is limited by Doppler broadening, since rovibrational transitions within the ground state are very weak and Doppler-free spectra are challenging to obtain in molecular beams.  The resulting Doppler-broadened spectra are difficult to interpret at the level of resolution beyond a few MHz \cite{CiuryloWcisloPRA16_DopplerLineshape}, while the current level of QED calculations calls for measurement precision at the kHz level or below.  This has motivated recent advances in Doppler-free methods (Sec. \ref{Sec:LD}), particularly with molecular hydrogen ions.

Light molecular ions are excellent candidates for testing QED and potentially measuring fundamental constants and their temporal stability.  For example, HD$^+$ ions in the ground electronic state were used to achieve a fractional resolution of $10^{-9}$ for rotational transitions in the Lamb-Dicke regime \cite{SchillerAlighanbariNPhys18_HDIonLambDicke}, as one of the earlier applications of this technique.  It was accomplished within a molecular ion crystal in a linear quadrupole trap with tight radial confinement, where the HD$^+$ ions were first sympathetically cooled with Be$^+$ atomic ions to $\sim10$ mK. In such ion crystals, the radial range of motion is typically $\sim10$ $\mu$m, whereas rotational wavelengths are $\sim0.1$-$1$ mm, such that the Lamb-Dicke condition is satisfied (Sec. \ref{Sec:LD}).

$Ab~initio$ calculations for ionic HD$^+$ {are more advanced} than for neutral molecular hydrogen. At several parts in $10^12$ \cite{KorobovPRL17_HMolIonsFewPPTCalc,GermannKoelemeijPRR21_HDIonQED5thForce}, these calculations are limited by the current precision of fundamental parameters, particularly $\eta$. Taking advantage of Lamb-Dicke spectroscopy in a Paul trap \cite{PatraKoelemeijScience20_HDIonProtonElectronRatioPPT} is therefore a promising path to extracting $\eta$ with a competitive uncertainty.  An ultrahigh precision of $3\times10^{-12}$ was achieved for the $(0,3)\rightarrow(9,3)$ optical rovibrational transition, where $(v,N)$ describes the vibrational and rotational quantum numbers.  The key was driving a two-photon transition with counter-propagating light beams with similar wavelengths, thus greatly suppressing the first-order Doppler shift, and subsequently counting the signal via state-selective photodissociation.  The natural linewidth is $\sim10$ Hz, although it was not resolved in this experiment.  The resulting precision for $\eta$ reached $2\times10^{-11}$, supporting the recent Penning-trap based measurements \cite{SturmHeissePRL017_ProtonMass}, and likely playing a role in future CODATA adjustments of fundamental constants (Fig. \ref{fig:qedmudot}(a)).  A different flavor of Lamb-Dicke spectroscopy was also used in trapped and sympathetically cooled HD$^+$ ensembles \cite{SchillerKortunovNPhys21_HDIonLambDickeMassRatio}.  Here, one-photon mid-infrared (5 $\mu$m) spectroscopy of the fundamental vibrational transition $(0,0)\rightarrow(1,1)$ circumvented Doppler broadening by confining sympathetically cooled ions in a string and interrogating them along an axis perpendicular to the trap.  This trapped ion cluster transverse excitation spectroscopy (
{Sec. \ref{Sec:LD}}) technique works for wavelengths that correspond to THz rovibrational transitions \cite{SchillerAlighanbariNature20_HDIonLambDickeQEDTests}.  The frequency uncertainty in this case also reached the ppt level, with a resulting $3.5\times10^{-11}$ uncertainty on $\eta$.

While rovibrational spectroscopy of HD$^+$ yields a competitive value of $\eta$, when combined with improved measurements of fundamental parameters \cite{SturmHeissePRL017_ProtonMass,BlaumRauNature20_DandHDIonMass}, antiprotonic helium spectroscopy ($\bar{p}^3$He and $\bar{p}^4$He), and new theoretical calculations that surpass predictions for any other molecular observable, it yields the tightest test of molecular QED \cite{GermannKoelemeijPRR21_HDIonQED5thForce}.
This was {accomplished} with a statistical approach designed to both test QED and set limits on new forces (Sec. \ref{sec:NewForces}). As a result, the theory of molecular vibrations and rotations was tested at the unprecedented levels of $4\times10^{-11}$ and $1\times10^{-10}$. Moreover, since the theoretical uncertainty for the vibrational transition $(0,3)\rightarrow(9,3)$ is largely due to QED contributions to the molecular energies, the analysis can be interpreted as a stringent test of molecular QED, below the $4\times10^{-6}$ level.  Gaining another order of magnitude is promising with further improvements in theory and fundamental constant measurements.  A notable feature is the presence of QED contributions that are unique to bound molecular systems and cannot be directly tested with atoms \cite{GermannKoelemeijPRR21_HDIonQED5thForce}.

\subsection{Temporal variations of fundamental constants \label{sec:temporalvar}}

\begin{figure*}[!]
\includegraphics[width=0.8\textwidth]{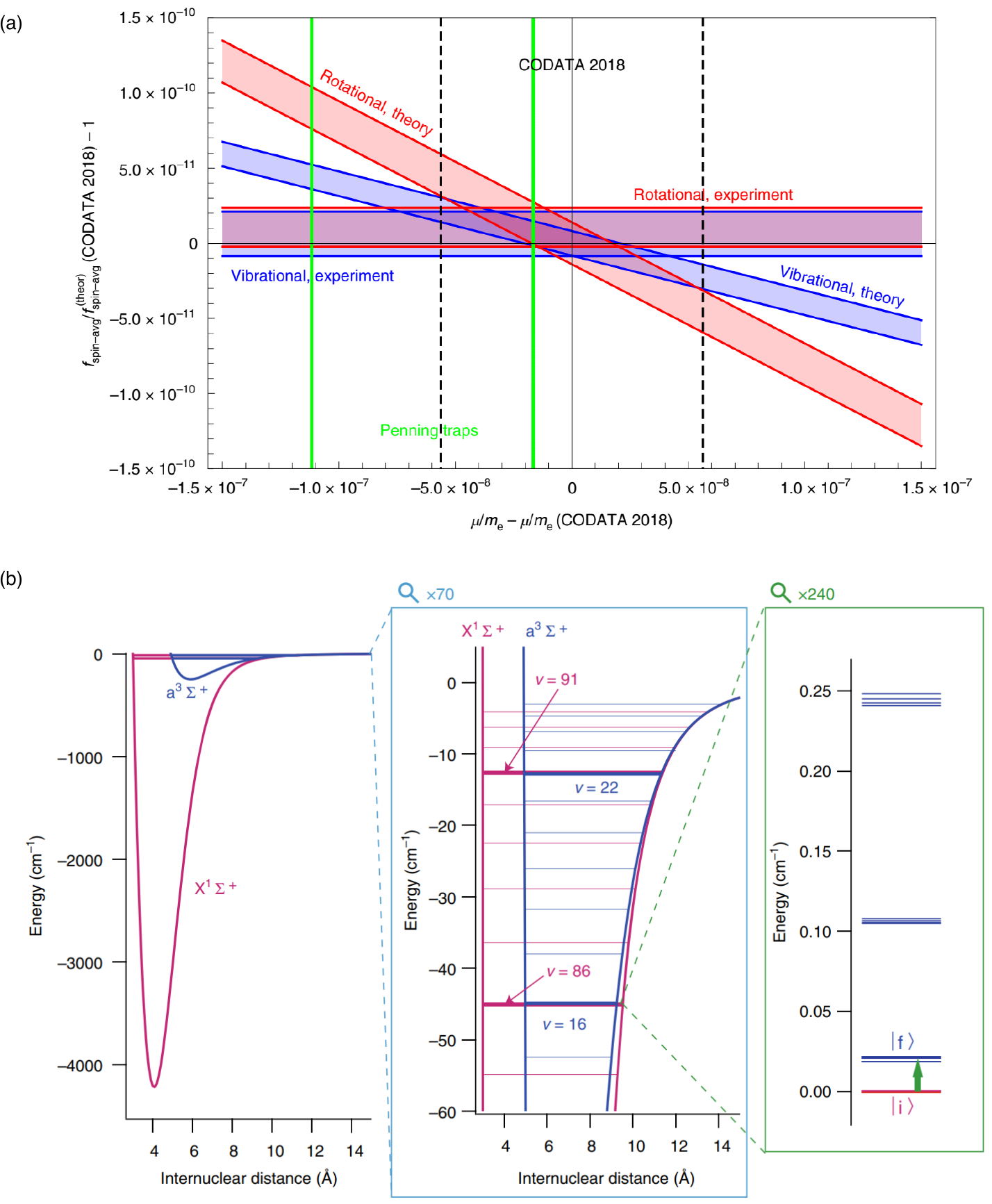}
\caption{\label{fig:qedmudot} The intimate connection between molecular spectra and the electron-to-proton mass ratio, $\eta$. (a) Determination of $\eta$ through Lamb-Dicke spectroscopy of the fundamental rovibrational overtone in HD$^+$, in excellent agreement with the recommended CODATA 2018 value and previous pure rotational spectroscopy of the same molecule. Adapted from \cite{SchillerKortunovNPhys21_HDIonLambDickeMassRatio}. (b) Bounds on a temporal drift of $\eta$ can be set by comparing the binding energies of rovibrational states with different sensitivity coefficients. Such a situation can arise between near-degenerate states of the singlet and triplet ground potentials of certain molecules, for example KRb as shown. Adapted from \cite{InouyeKobayashiNComm19_MuDotKRb}.}
\end{figure*}

Molecular ions are intriguing candidates for optical clocks \cite{KorobovSchillerPRL14_H2IonClocks} and sensors of possible variations in fundamental constants.  Such variations would be a signature of new physics where the laws of the Universe depend on time or position.  Molecules are particularly sensitive to changes in the electron-to-proton mass ratio, $\dot{\eta}$ \cite{HannekeQST21_MolecularClocks,HannekeCarolloAtoms19_O2IonMassRatioVar,HannekePRA16_O2Ion,OdomKokishPRA18_MuDotWithMolIons}.  In fact, $\eta$ is sensitive to a wide range of new interactions that may couple the quantum chromodynamics energy scale, $\Lambda_{\mathrm{QCD}}$ which determines nucleon masses, to the Higgs field which gives rise to the electron mass.  Generally, unification models predict a larger fractional variation in $\eta$ than in the fine structure constant $\alpha$, potentially resulting in an easier detection of $\dot{\eta}$ \cite{LangackerPLB02_UnificationAlphaDot,FlambaumPRD04_AlphaQCDVariations}.  The $\eta$ variations could consist of oscillations or slow drifts \cite{HannekeQST21_MolecularClocks}.  Beyond-Standard-Model physics that connects fundamental constants to cosmological evolution and extra spatial dimensions typically implies slow drifts of the constants \cite{UzanRMP03}.  On the other hand, models that involve new massive particles and associated fields often imply oscillatory behavior of fundamental constants \cite{FlambaumStadnikPRL15_ConstantVariationsDarkMatter,VanTilburgArvanitakiPRD15_DilatonsAtomicClocks,KozyryevDoylPRA2021_enhancedSrOH_mudot}.  This can occur if the new particles, which could constitute dark matter, possess a sufficiently low mass and high density, and therefore feature long, overlapping Compton wavelengths.  Particles of this type would oscillate coherently as a wave, and their couplings to nuclei and electrons could then induce oscillations of fundamental constants. The range of new boson masses allowed by such models covers $\sim20$ orders of magnitude, leading to a widely open parameter space of possible oscillation frequencies.  To date, experiments have largely focused on changes in fundamental constants that occur on a slower time scale than the experimental cycle.
 
Molecules including H$_2$ have been used to measure the drift of $\eta$ that might have occurred since the times of the early Universe $\sim10^{10}$ years ago \cite{SafronovaRMP18_NewPhysicsAtomsMolecules}.  These measurements rely on molecules that are relevant to generating quasar absorption lines, and take advantage of the large time differential.  In contrast, measurements of the present-day $\dot{\eta}$ have the advantage of engineering an extraordinary clock-like level of precision at the expense of the relatively short measurement duration of $\sim1$ year.  Astronomical observations of methanol spectra yield $\dot{\eta}/\eta\lesssim2\times10^{-17}/$year over billions of years \cite{UbachsBagdonaitePRL13_MuDotMethanol}.  The most stringent present-day laboratory-based test to date relies on optical (Yb$^+$) and microwave (Cs) atomic clocks, and reveals that $\dot{\eta}/\eta\lesssim4\times10^{-17}/$year \cite{PeikLangePRL21_AlphaDotMuDot}.  Here the sensitivity to $\dot{\eta}$ arises from the nuclear magnetic moment of Cs and is therefore model-dependent.  On the other hand, molecules are directly sensitive to $\dot{\eta}$ because their vibrational and rotational spectra depend on both nuclear and electron masses. Vibrational molecular transitions are particularly attractive for present-day $\dot{\eta}$ measurements.  The best such measurement to date constrains $\dot{\eta}/\eta$ to $\sim1\times10^{-14}/$year and is based on vibrational states in ultracold KRb \cite{InouyeKobayashiNComm19_MuDotKRb} (Fig.~\ref{fig:qedmudot}(b)), superseding an early SF$_6$ beam-based result \cite{ChardonnetShelkovnikovPRL08_muStability}.

A high degree of quantum-state control, such as that recently achieved for ultracold Sr$_2$ molecules \cite{ZelevinskyPRL08,ZelevinskyKondovNPhys19_MolecularClock}, is essential for high-precision experiments.  However, molecules that can be assembled from laser-cooled atoms have relatively small binding energies, while the intrinsic sensitivity to $\dot{\eta}$ is enhanced for deeply bound molecules \cite{HannekeQST21_MolecularClocks,ZelevinskyPRL08}.  Ideally, these are homonuclear diatomic molecules with dipole-forbidden vibrational transitions which can be driven via the electric quadrupole or two-photon mechanism. The fractional sensitivity to changes in $\eta$ scales as the fractional frequency sensitivity of the experiment, divided by the relative sensitivity factor $K_{\eta}$, so that $\delta\eta/\eta=(\delta f/f)/K_{\eta}$.  For vibrational transitions $K_{\eta}=-0.5$, potentially allowing the statistical sensitivities to reach the best current limit \cite{PeikLangePRL21_AlphaDotMuDot} within hours of averaging:  systematic uncertainties on single-ion atomic clocks that use similar techniques to the molecular-ion based $\dot{\eta}$ measurement proposals have recently been reported below $10^{-18}$ \cite{LeibrandtBrewerPRL19_AlIonClock}.

One attraction of molecules in measuring changes in fundamental constants is that a single molecular species can offer a range of transitions with different sensitivities to $\dot{\eta}$, but which are susceptible largely to the same systematic shifts, thus improving the simplicity and accuracy of experiments.  Large binding energies and a high degree of control can be achieved by working with trapped molecular ions of homonuclear diatomics, especially molecules such as $^{16}$O$_2^+$ that are free of hyperfine structure \cite{HannekePRA16_O2Ion}.  In particular, for O$_2^+$ it is possible to achieve $\sim50$-fold improved sensitivities compared to typical ultracold molecules \cite{InouyeKobayashiNComm19_MuDotKRb,ZelevinskyPRL08} by driving two-photon transitions between the most deeply bound vibrational states and the intermediate-bound states ($v\sim30$) \cite{HannekeQST21_MolecularClocks}.  Alternatively, $\sim20$-fold improved sensitivities can be reached by driving direct electric-quadrupole transitions to $v\sim5$-$10$ \cite{KajitaPRA17_O2IonMuDot}. Homonuclear molecules also {diminish} the effects of DC Stark shifts and blackbody radiation on the transition frequencies, improving the outlook for systematic shifts (Sec. \ref{Sec:VibCoh}).

Nonpolar homonuclear molecules have distinct advantages for precision metrology such as $\dot{\eta}$ measurements, yet certain polar molecules offer a different set of benefits.  In particular, TeH$^+$ ions have been explored for such experiments \cite{OdomKokishPRA18_MuDotWithMolIons,OdomStollenwerkAtoms18_TeHIonOpticalPumping}.  The vibrational clock transitions here offer high sensitivities as for O$_2^+$, but are dipole-allowed.  Importantly, this ion belongs to a class of molecules with highly diagonal Franck-Condon factors, implying a possibility to cycle multiple photons in order to implement efficient state preparation or high-fidelity detection protocols.  In polar molecules, DC and AC Stark shifts could potentially limit precision through poorly controlled systematic shifts.  However, the scalar polarizability shifts can be small, while tensor polarizabilities could be suppressed through state averaging techniques \cite{OdomKokishPRA18_MuDotWithMolIons}.

While there is currently a gap of two to three orders of magnitude between atomic and molecular measurements of $\dot{\eta}$, it is expected to be closed with a prudent choice of molecules that allow stringent control of systematic and statistical uncertainties.  In the coming decade, we can expect molecular clocks to increasingly contribute to precision measurements of fundamental constant variations, particularly for long-term drifts and slow oscillations, while deepening our understanding of the effects that limit fundamental metrology with molecules \cite{ZelevinskyKondovNPhys19_MolecularClock}.  Generally, molecular clocks based on vibrations and rotations should be largely analogous to those encountered with electronic transitions used in atomic clocks \cite{SchmidtLudlowRMP15_OpticalAtomicClocks}.  The key challenges of working with molecules can involve lower overall transition energies and thus longer coherence times necessary to obtain a matching frequency resolution, as well as less efficient methods of producing samples of neutral state-controlled molecules.  The advantages of vibrational clocks include potentially common-mode systematic effects, since clock states have the same electronic quantum numbers.  The main systematic effects that will affect molecular clock measurements are related to Stark shifts from DC and AC electric fields.  Polar molecules are subject to relatively large DC Stark shifts, although these potentially can be canceled by a clever choice of transition \cite{OdomKokishPRA18_MuDotWithMolIons}. The laser-induced AC Stark shifts can be troublesome for vibrational transitions that cannot be driven directly and thus require two-photon schemes \cite{ZelevinskyKondovNPhys19_MolecularClock}.  For optical-lattice-trapped neutral molecules, the development of magic trapping is key to suppressing trap-induced Stark shifts (Sec. \ref{Sec:VibCoh}).  Finally, second-order Doppler shifts that currently limit optical atomic ion clocks \cite{LeibrandtBrewerPRL19_AlIonClock} are also expected to be a dominant uncertainty for molecular {ion} clocks, although at a low level of $<10^{-18}$.

\subsection{New physics with molecular spectroscopy}
\label{sec:NewForces}

\begin{figure}[!]
\includegraphics[width=\textwidth]{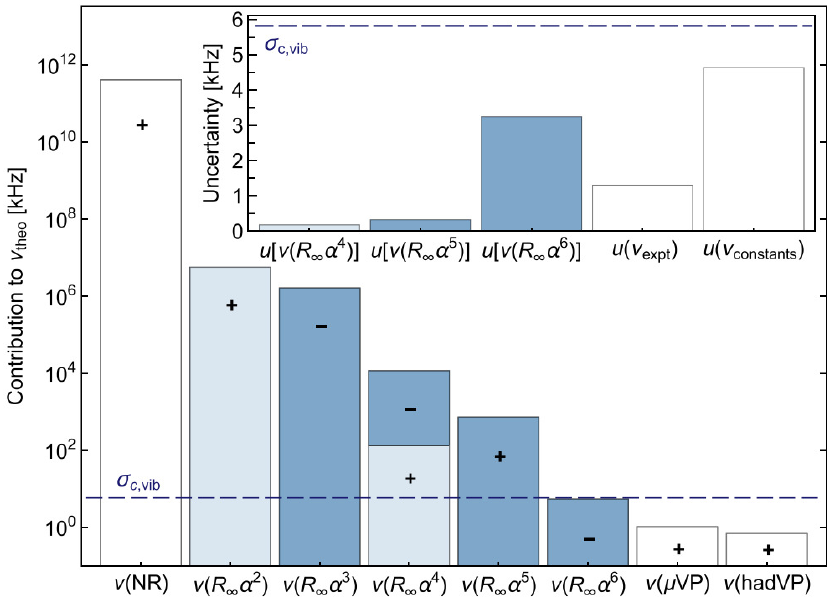}
\caption{\label{fig:fifth} Contributions of various terms to an $ab~initio$ calculation for the $(0, 3) \rightarrow (9, 3)$ transition frequency in HD$^+$. Simple molecules like HD$^+$ are especially amenable to quantum chemistry modeling. Comparison with {experimental} measurements placed bounds on new scalar hadron-hadron interactions.  {Relativistic terms are highlighted in blue, with dark blue denoting QED corrections.  The sign of each contribution is indicated, and the dashed line is the combined uncertainty.  The uncertainty sources are shown in the inset, illustrating that the experimental error is small enough to test the limit of three-body QED calculations.}  Adapted from \cite{GermannKoelemeijPRR21_HDIonQED5thForce}.}
\end{figure}

Table-top experiments in atomic physics \cite{MullerParkerScience18_Alpha,GuellatiKhelifaMorelNature20_AlphaRb,UdemGrininScience20_H1S3S} have opened a window on possible new physics beyond the Standard Model.  This is possible by virtue of their ultrahigh precision, recently reaching the range of parts per trillion for several determinations of fundamental constants, combined with the success of theoretical QED.  This level of precision, both theoretical and experimental, has eluded even the simplest of molecules until recently.  Diatomic molecules have now entered the scene and can illuminate possible new physics that is complementary to what atomic systems probe.  In particular, in a diatomic molecule the relatively massive pair of nuclei can sense gravity-like, or mass-dependent, new physics \cite{FayetCQG96_NewInteractions}, including long-range nucleon-nucleon ``fifth" forces to which atomic systems are insensitive.  In molecular spectroscopy that samples molecular vibrations and rotations, transition frequencies directly depend on nuclear masses.
Molecular spectroscopy can be also interpreted in terms of constraints on compactification distances \cite{UbachsSalumbidesNJP15_ExtraDimensionsMolecSpectrosc} for theories that explain the fundamental hierarchy problem by postulating additional compactified dimensions for gravity \cite{ArkaniHamedPLB98_HierarchyProblemMMDimensions}.

Newtonian gravity is well established at large distances down to the micron scale, but it is far less well tested at shorter distances because electromagnetic forces overwhelm gravity in this regime.  Such tests predominantly originate from neutron scattering \cite{KomamiyaKamiyaPRL15_NmGravityConstraintNeutrons,YoshiokaHaddockPRD18_NmGravityNeutronsImproved} and Casimir-force type experiments \cite{MostepanenkoBordagPhysRep01_CasimirDevelopments}.  Possible corrections to the Newtonian gravitational potential or new nucleon-nucleon interactions can be parametrized, for example, as a Yukawa potential $\beta e^{-R/\lambda}/R$, where $R$ is the separation between the masses, $\lambda$ is the length scale of the new force, and $\beta$ gives the interaction strength.  If the new force arises from light bosons of mass $M$ that couple to nucleons, then $\lambda=\hbar/(Mc)$ where $c$ is the speed of light.  Ultrahigh-precision rovibrational spectroscopy of molecules is an exciting direction that utilizes fully quantum systems to extend the search for new physical forces \cite{UbachsSalumbidesPRD13_FifthForcesMolecPrecisMeasts,TakahashiBorkowskiSR19_Yb2Gravity}. Here, $R$ can be naturally varied by utilizing states with different binding energies.

As discussed in Sec. \ref{sec:QEDtests}, spectroscopy of molecular hydrogen isotopologues H$_2$, D$_2$, and HD has enabled high-precision tests of few-body QED.  In particular, rovibrational spectroscopy of HD$^+$ ions in the Lamb-Dicke regime has led not only to the most rigorous molecular QED test at the parts-per-million level, but also to a tightened constraint on forces beyond the Standard Model at the angstrom scale that corresponds to the bond length in hydrogen molecules \cite{GermannKoelemeijPRR21_HDIonQED5thForce}.  The specific HD$^+$ transitions that contribute to state-of-the-art fifth-force constraints are $(0,3)\rightarrow(9,3)$ \cite{PatraKoelemeijScience20_HDIonProtonElectronRatioPPT}, $(0,0)\rightarrow(0,1)$ \cite{SchillerAlighanbariNature20_HDIonLambDickeQEDTests}, and $(0,0)\rightarrow(1,1)$ \cite{SchillerKortunovNPhys21_HDIonLambDickeMassRatio}, where the transitions cover the frequency range from 1 to 400 THz (Fig. \ref{fig:fifth}).  Combining these measurements with improved $ab~initio$ calculations, a statistical analysis sets limits on new forces of the Yukawa type \cite{GermannKoelemeijPRR21_HDIonQED5thForce}.  Molecules are particularly sensitive to nucleon-nucleon forces, and the molecule-based constraint is now within two orders of magnitude of the neutron-scattering result which {depends on nuclear parameters and collision models} \cite{KomamiyaKamiyaPRL15_NmGravityConstraintNeutrons,YoshiokaHaddockPRD18_NmGravityNeutronsImproved}.  In contrast, HD$^+$ based constraints rely only on relativistic quantum mechanics and QED.  At the angstrom length scale, the fifth force is now excluded to approximately $10^{-11}$ of the strength of the electromagnetic interaction given by the fine structure constant.  The hydrogen molecule system is also competitive for studying new electron-neutron forces, which can be constrained via atomic isotope shifts \cite{VuleticCountsPRL20_YbIonKingPlot,DrewsenSolaroPRL20_CaIonKingPlot,SoreqDelaunayPRD17_LightAtomsSpinDepForces}.  However, these results depend on the Rydberg constant which itself could be affected by electron-nucleon coupling.

Given the recent access to trapped ultracold molecular samples, it is tempting to extend searches for fifth forces to molecules amenable to excellent quantum control.  This is the case despite the additional theoretical challenges presented by the generally heavier molecules that are available at ultracold temperatures.  Diatomic molecules based on alkaline-earth-metal atoms are particularly attractive since they possess narrow optical transitions (see also Sec. \ref{sec:forbid}) that facilitate their creation, control, and detection \cite{ZelevinskyPRL06,ZelevinskyMcGuyerNPhys15_Sr2M1,ZelevinskyMcDonaldPRL15_Sr2LatticeThermometry,TakahashiTojoPRL06_YbNarrowLinePA,TakahashiTakasuPRL12_Yb2Subradiant}.  Importantly, bosonic isotopes of these atoms have no electronic or nuclear spin in the ground state.  This feature leads to purely van der Waals molecular ground potentials and greatly simplifies the theoretical interpretation of vibrational bound states.  As for hydrogen molecules (Sec. \ref{sec:QEDtests}), $ab~initio$ theoretical descriptions of alkaline-earth-metal dimers begin with the Born-Oppenheimer approximation followed by corrections that include adiabatic and nonadiabatic effects, nuclear-size effects, and relativistic and QED corrections.

The dimers of Yb have been explored fairly extensively in the context of improving the understanding of mass-dependent interatomic forces, partly because there are multiple spinless isotopes.  Yb$_2$ molecules are not currently  available in deeply bound vibrational states, so existing work focuses on the near-threshold states ($R\sim5$ nm) that can be accessed via photoassociation spectroscopy \cite{JonesRMP06}.  The binding energies have been determined at sub-kHz precision for several isotopologues \cite{TakahashiBorkowskiPRA17_YbPABeyondBO}, and the long-range interatomic interaction was theoretically modeled with a partially $ab~initio$ Born-Oppenheimer potential as well as by including some mass-dependent beyond-Born-Oppenheimer corrections \cite{HutsonLutzJMS16_BODeviations}, both adiabatic and nonadiabatic, which significantly improved agreement with the data.  $Ab~initio$ calculations are not directly competitive with those for molecular hydrogen; on the other hand, mass-dependent corrections are fractionally smaller for heavy molecules, and a selection of different masses is available for experiments.  Further improving the theoretical precision, especially facilitated by using molecules with lighter alkaline-earth-metal constituents, and probing a wider range of internuclear separations is promising for setting competitive limits on new forces in the nanometer range.

\subsection{Highly forbidden transitions}
\label{sec:forbid}

A transition between two states is said to be forbidden when angular momentum selection rules dictate that the probability for an electric dipole (E1) transition is zero. Nevertheless, in most cases, higher-order coupling beyond the lowest-order electric dipole can still occur (e.g., magnetic dipole M1, electric quadrupole E2, etc.). 
Forbidden transitions are central to high-performance quantum clocks. One attraction of building a molecular clock is the existence of multiple clock transitions within the same system simply by changing the choice of $v$ or $J$, which, when interrogated in an interleaved fashion, allows for convenient common-mode rejection of systematic effects (see also Sec. \ref{sec:temporalvar}).

Establishment of a link between optical frequencies and the microwave domain for frequency counting has been made possible by the advent of octave spanning optical frequency combs. Comparison {of a molecular clock} to the existing global network of clocks allows for tests of the variation of fundamental constants \cite{BarontiniQSNET21} while demonstrating control of light-matter interactions in a frequency range spanning over six orders of magnitude: from microwave rotational or hyperfine transitions to infrared vibrational transitions to optical electronic transitions. To list a few examples, an optical molecular clock based on the 3.39 $\mu$m $F_2^{(2)}$ transition in CH$_4$ reached a fractional instability of $1.2 \times 10^{-13}$ in one second \cite{foreman2005demonstration}. A molecular fountain was shown for NH$_3$ where the molecules were in free fall for $\sim$300 ms, and the same principles could be extended to build fountain clocks with sub-Hz Ramsey spectroscopy \cite{UbachsChengPRL16_molecularfountain}. A recent proposal suggests that a 3D lattice {(or tweezer)} clock could be realized with laser-cooled molecules such as CaF, where the $Q(0,a)$, $Q(0,b)$, and $Q(1)$ forbidden transitions at $\sim$17 THz can be accessed using a pair of Raman lasers \cite{kajita2018precise,BarontiniQSNET21}. 

Neutral and ionic hydrogen are simple yet interesting systems for observing highly forbidden transitions. For example, transitions between nuclear spin configurations of a molecule (ortho-para transitions) can occur only via `gerade-ungerade' mixing induced by nuclear hyperfine interactions. Since hyperfine couplings tend to be small, ortho-para transitions are typically too weak to be observed but their transition strengths can be magnified near the dissociation threshold where the density of states is high.  Indeed an ortho-para transition was observed in H$_2^+$ molecules by looking at a pure rotational line ($v=19$, $\Delta v = 0$) near the dissociation threshold, with a precision of $\sim1$ MHz and consistent with predictions \cite{McNabCritchleyPRL01_H2IonOrthoParaTransition}.  In neutral H$_2$, the difference between the energy level structures of the two nuclear spin isomers was constrained to a similar precision \cite{MerktBeyerPRL19_H2ParaOrthoInterval}. In another experiment, radiative ortho-para transitions with linewidths of $<5$ kHz were observed in S$_2$Cl$_2$ via Fourier transform microwave spectroscopy \cite{EndoKanamoriPRL17_S2Cl2OrthoPara}. This was facilitated by the unique structure of the S$_2$Cl$_2$ molecule:  its helically twisted geometry leads to strong coupling between its nuclear quadrupole moments and the internal electric field gradient, and being a prolate asymmetric top it possesses numerous near degenerate levels with opposite ortho-para symmetry. Electric quadrupole transitions in molecular ions were first observed in N$_2^+$ for both its $I=0,2$ nuclear spin isomers \cite{WillitschGermannNatPhys14_dipoleforbidden_N2}. The rotational components of the forbidden $S(0)$ line ($v=0\rightarrow 1$, $J=0\rightarrow 2$) were measured with linewidths of $\sim$20 MHz limited by the instability of the quantum cascade laser driving the transitions.

Homonuclear molecules allow for the study of phenomena such as two-body ``subradiance" and ``superradiance''. In the former, destructive interference of neighboring light scatterers lead to a suppression of spontaneous emission rates, as was pointed out by Dicke \cite{DickePR54_CollectiveRadiation}. Optical transitions in homonuclear diatomic molecules with wavelengths greater than the typical interatomic separation realize the simplest configuration of the Dicke model. {Due to strong} subradiance {E1 suppression}, the M1 and E2 transitions $X0_g^+ \rightarrow 1_g$ near the ${^1S}_0+{^3P}_1$ asymptote have been observed in ultracold Sr$_2$ \cite{ZelevinskyMcGuyerNPhys15_Sr2M1} and Yb$_2$ \cite{TakahashiTakasuPRL12_Yb2Subradiant} with narrow linewidths of $\sim$0.1--1 kHz limited by predissociation to the energetically lower ${^1S}_0+{^3P}_0$ threshold. 

Forbidden transitions can gain intensity through the application of an external field. For example, it has been proposed that weakly bound molecules of closed-shell atoms (Sr$_2$, Yb$_2$) can serve as precise optical molecular clocks \cite{BorkowskiPRL18}. Here, by analogy with optical atomic clocks of the same species, the doubly forbidden transition from ground state molecules to the excited $0_u^-$ (asymptoting to ${^1S}_0+{^3P}_0$) can be induced by an external magnetic field by admixing with the ${^3P}_1$ state \cite{OatesTaichenachevPRL06}. Magnetic field control over the strength of normally forbidden $\Delta J = 2,3$ molecular transitions was investigated in Sr$_2$ \cite{ZelevinskyMcGuyerPRL15_Sr2ForbiddenE1}. The dense structure near the ${^1S}_0+{^3P}_1$ threshold results in characteristic mixing fields of only several tens of gauss, a factor of nearly $1\times10^6$ less than the fields typically required for atoms.  {In general, the wealth and possible tunability of forbidden transitions in molecules open the door to exciting applications in metrology and high-precision measurements.}

\section{Related future directions}
\label{Sec:Future}

Besides the approaches and efforts described above, there are many {possible} paths {toward} probing fundamental physics phenomena with molecules.  For example, ultracold van der Waals dimers can enrich the chapter on nm-scale mass-dependent interatomic force measurements due to the highly refined quantum control they can afford, as described in Sec. \ref{sec:NewForces}.  These molecules do not form strong chemical bonds, unlike H$_2$, and probe slightly longer spatial scales of $R\sim0.5$-$5$ nm.  So far,
weakly bound Yb$_2$ molecules have been subject of such experiments \cite{TakahashiBorkowskiPRA17_YbPABeyondBO,TakahashiBorkowskiSR19_Yb2Gravity}.  However, theoretical calculations are extremely challenging for heavy atomic and molecular species.
The lighter Sr$_2$ molecules have a notably simpler constituent atomic structure.  These molecules present an exciting prospect for experiments on ultrashort-range forces.  They are photoassociated from Sr atoms that are easily laser-cooled to just $\sim1$ $\mu$K \cite{ZelevinskyReinaudiPRL12_Sr2}.  Moreover, these molecules have been efficiently created in weakly bound and deeply bound vibrational states \cite{ZelevinskyLeungNJP21_STIRAP}, and are amenable to atomic-clock level metrological precision \cite{ZelevinskyKondovNPhys19_MolecularClock,ZelevinskyLeungPRL20_MClockRabi100ms}.
Ultrahigh-precision vibrational spectra of these molecules can be initially studied via King plot analysis methods in combination with state-of-the-art theory that is {actively} under development.  {Since many vibrational quantum states are accessible -- covering a broad range of $R$ -- new forces may be easier to distinguish from QED terms via their different asymptotic dependence on $R$.}

In Sec. \ref{Sec:FundSym} we discussed the use of state-controlled molecules for searches of parity symmetry ($P$) violation that, in conjunction with $T$ violation, would herald physics beyond the Standard Model.  However, the fundamental weak force is known to cause $P$ violation that is consistent with SM, but difficult to detect due to its small size \cite{BouchiatROPP97_AtomsPV}.  Measuring the effects of the weak force with an improved precision will help test the predictions of the SM weak sector, and to potentially discover new physics beyond the SM.  In particular, an interesting effect to improve upon is the product of two types of weak force charges:  the axial{-vector} nuclear charge and the {polar-vector} electronic charge.  It turns out that molecules can amplify not only the existence of $P$ violation via permanent EDMs (Sec. \ref{Sec:FundSym}) but also the effects of electroweak $P$ violation, as compared to atomic systems.  The underlying principle is to expose the molecules to electric and magnetic fields of different ``handedness" and then detect the differential
energy shift.  Such experiments are underway for BaF molecules, with fermionic Ba isotopes that possess nuclear spin \cite{DeMilleAltunasPRL18_NSDPVMethod}.  Nuclear-spin-dependent parity violation (NSD-PV) is generally weaker than spin-independent parity violation and challenging to measure precisely.  This effect mixes opposite-parity energy levels, and is therefore magnified in molecules with nearby levels of opposite parity which can be brought even closer together with magnetic fields \cite{DeMilleCahnPRL14_BaFLevelCrossing}.  We note that BaF molecules are laser coolable, which in the future can yield a significant enhancement of NSD-PV signals \cite{LangenKogelNJP21_BaFCoolingScheme}.  Furthermore, if diatomic molecules are promising for NSD-PV, polyatomic molecules can offer even larger gains.  For example, laser-coolable triatomics such as BaOH, YbOH, and RaOH feature closely spaced $l$-doublets (Sec. \ref{SubSec:direct_cooling}) which can allow the opposite-parity states to reach degeneracy at 100-fold lower magnetic fields \cite{NorrgardCP2019_PolyMolParity}, promising simpler experiments that are less prone to systematic shifts.

\section{Conclusion}
\label{Sec:Conclusion}

{The multitude of research directions discussed in this Perspective are examples of new ideas that are driving the field of precision measurements with molecules.  This field is growing rapidly because of a confluence of two major factors.  One factor is the great improvement in molecular quantum control in recent years, which has been motivated by advancements in quantum information science.  The second factor is the apparent incompleteness of the Standard Model of physics, which is difficult and costly to directly address with existing high-energy techniques. This dual motivation, and the early success of the field, makes us confident that in the coming years we will witness an abundance of new results and insights gained from the molecular approach to high-precision measurements in fundamental physics.}

\end{document}